\documentclass[pdflatex,sn-mathphys-num]{sn-jnl}

\usepackage[table]{xcolor}
\usepackage{graphicx}%
\usepackage{multirow}%
\usepackage{amsmath,amssymb,amsfonts}%
\usepackage{amsthm}%
\usepackage{mathrsfs}%
\usepackage[title]{appendix}%
\usepackage{xcolor}%
\usepackage{textcomp}%
\usepackage{manyfoot}%
\usepackage{booktabs}%
\usepackage{algorithm}%
\usepackage{algorithmicx}%
\usepackage{algpseudocode}%
\usepackage{listings}%
\usepackage{makecell}%
\usepackage{caption}%
\usepackage{bm}

\theoremstyle{thmstyleone}%
\theoremstyle{thmstyletwo}%

\theoremstyle{thmstylethree}%

\begin{document}

\title[Seeing the Invisible through Speckle Images]{Seeing the Invisible through Speckle Images}


\author*[1]{\fnm{Weiru} \sur{Fan}}\email{weiru\_fan@zju.edu.cn}
\equalcont{These authors contributed equally to this work.}

\author[1]{\fnm{Xiaobin} \sur{Tang}}\email{12036056@zju.edu.cn}
\equalcont{These authors contributed equally to this work.}

\author[1]{\fnm{Xingqi} \sur{Xu}}\email{xuxingqi@zju.edu.cn}

\author[2]{\fnm{Huizhu} \sur{Hu}}\email{huhuizhu2000@zju.edu.cn}

\author[3]{\fnm{Vladislav V.} \sur{Yakovlev}} \email{yakovlev@tamu.edu}

\author[1]{\fnm{Shi-Yao} \sur{Zhu}}\email{syzhu@zju.edu.cn}

\author*[1,2,4]{\fnm{Da-Wei} \sur{Wang}}\email{dwwang@zju.edu.cn}

\author*[1]{\fnm{Delong} \sur{Zhang}}\email{dlzhang@zju.edu.cn}

\affil*[1]{\orgdiv{Zhejiang Key Laboratory of Micro-Nano Quantum Chips and Quantum Control, School of Physics, and State Key Laboratory for Extreme Photonics and Instrumentation}, \orgname{Zhejiang University}, \orgaddress{\city{Hangzhou}, \postcode{310027}, \state{Zhejiang Province}, \country{China}}}

\affil[2]{\orgdiv{College of Optical Science and Engineering}, \orgname{Zhejiang University}, \orgaddress{\city{Hangzhou}, \postcode{310027}, \state{Zhejiang Province}, \country{China}}}

\affil[3]{\orgdiv{Department of Biomedical Engineering}, \orgname{Texas A\&M University}, \orgaddress{\city{College Station}, \postcode{77843}, \state{TX}, \country{USA}}}

\affil[4]{\orgdiv{Hefei National Laboratory}, \orgaddress{\city{Hefei}, \postcode{230088}, \state{Anhui province}, \country{China}}}


\abstract{Scattering obscures information carried by wave by producing a speckle pattern, posing a common challenge across various fields, including microscopy and astronomy. Traditional methods for extracting information from speckles often rely on significant physical assumptions, complex devices, or intricate algorithms. Recently, machine learning has emerged as a scalable and widely adopted tool for interpreting speckle patterns. However, most current machine learning techniques depend heavily on supervised training with extensive labeled datasets, which is problematic when labels are unavailable. To address this, we propose a strategy based on unsupervised learning for speckle recognition and evaluation, enabling to capture high-level information, such as object classes, directly from speckles without labeled data. By deriving invariant features from speckles, this method allows for the classification of speckles and facilitates diverse applications in image sensing. We experimentally validated our strategy through two significant applications: a noninvasive glucose monitoring system capable of differentiating time-lapse glucose concentrations, and a high-throughput communication system utilizing multimode fibers in dynamic environments. The versatility of this method holds promise for a broad range of far-reaching applications, including biomedical diagnostics, quantum network decoupling, and remote sensing.}




\maketitle

\section{Introduction}\label{sec1}
When coherent waves pass through inhomogeneous media, elastic light scattering leads to random wave interference producing speckle patterns characterized by a variable distribution of intensities \cite{rotter2017light, savo2017observation}. This effect often disrupts the information carried by the incident wave, hindering the imaging and sensing capabilities of optical systems. For many applications involving coherent waves, such as optical and ultrasound imaging through turbid media, speckle is often considered noise, and sophisticated signal/image processing methods are applied to minimize speckles \cite{redding2012speckle, bianco2018strategies, michailovich2006despeckling}. However, speckle patterns also contain important structural information, and their analysis can lead to achieving super-resolution imaging \cite{mudry2012structured, choi2022wide} and provide texture assessment for different samples \cite{gossage2003texture}. Accurate decoding of these speckle patterns might reveal a wealth of information to allow for better imaging and sensing. A stunning example of utilizing speckles for high-resolution spectroscopic measurements has recently been demonstrated by Cao’s group at Yale University \cite{redding2013compact}. Extensive computational analysis of speckles can also facilitate lensless 3D single-shot imaging through scattering media \cite{antipa2017diffusercam}.\par

To take full advantage of the wealth of information encoded in speckle patterns, statistical approaches based on correlation analysis \cite{bertolotti2012non, katz2014non} and transmission matrix \cite{popoff2010measuring, fan2021high} have been employed. Such strategies have resulted in a remarkable success of speckle imaging for biomedical applications \cite{silva2022signal}. However, such speckle analysis strongly depends on the sample’s structure and texture due to the fact that different tissues have a different size distribution of scatterers, making it difficult to generalize this approach to arbitrary scattering media. Machine learning (ML), on the other hand, provides a powerful tool to process complex data, showing promise in addressing the challenges of speckle interpretation and imaging while possessing high generalizability and scalability \cite{li2018deep, rahmani2018multimode, li2018imaging, borhani2018learning}. To this point, computational methods based on supervised learning, however, have been dominated by image reconstruction from speckles which requires an extensive library of ground truth measurements which may or may not be available for the system of interest. Furthermore, image sensing, which can be described in terms of the position or shape or any other specific property of an object under study, does not necessarily require the whole image to be reconstructed with great accuracy because it commonly contains a massive amount of redundant or irrelevant information \cite{wang2023image}. A ground-breaking paradigm of image sensing utilizes scattering as a means of encoding and then directly extract the salient features to compress images into a low-dimensional space. In this framework, the optical system and related algorithms do not utilize speckle patterns for imaging purposes, but search for high-level information for the downstream tasks. This is conceptually similar to an approach widely used in biomedical imaging in utilizing dynamic speckle information to assess blood flow \cite{boas2010laser}.\par

In general, the concept of "class" represents high-level information that is essential for understanding and managing complex data, such as speckles. For instance, the proper classification of speckles can be used to aid in diagnosing patients' health conditions \cite{wang2021deep, sezer2020deep}, while different classes of speckles in optical communication may represent transmitted signals or data values \cite{gong2019optical, zhu2021compensation}. Despite the overall simplification of the initial problem, the accurate speckle classification remain enormously challenging when using standard image processing methods, as even a minor shit in the incident wavefront can lead to substantial variations in speckle patterns. ML has been proposed as a way to implement a high-performance classifier and has also demonstrated its scalability for characterizing speckle patterns and for imaging from speckles \cite{li2018deep, luo2022computational}. In one of such illustrative examples, ML was successfully employed to learn the transmissive properties of scattering media \cite{li2018deep}. However, the ultimate performance of ML significantly relies on the design of the computational graph and training \cite{lecun2015deep}. To perform speckle interpretation, an enormous amount of labeled data is required to successfully calibrate the computational graph. Unfortunately, in most situations, the ground truth is often unknown making it impossible to obtain labeled data and to train; thus, a lack of sufficient ground truth for training results in limited generalization \cite{poggio2020theoretical}. To mitigate this problem, transfer learning and semi-supervised learning are proposed to reduce the requirement of labeled data for training \cite{zhuang2020comprehensive, van2020survey, rahmani2020actor}. However, even a small amount of labeled data is still unavailable in many practical applications, especially involving never-before-seen scattering media. Therefore, there is a clear need to explore alternative approaches.\par

Here, we introduce and validate a new paradigm of speckle sensing through the development of an automated clustering algorithm for speckle unsupervised recognition and evaluation (SURE). One of the most attractive features of SURE is that it does not require the ground truth or labels for supervised training; instead, it only utilizes input speckles to explore internal relations hidden among them (see Tab. \ref{tab1} for properties of SURE). The central idea of SURE is to cluster data by bootstrapping the high-level features among different speckles and then assigns corresponding labels to the cluster groups according to the preset protocol. As a demonstration and validation of this novel approach, we show in this report that SURE takes advantage of light scattering to encode information rather than viewing it as an obstacle, enabling noninvasive glucose monitoring. Such a scheme allows for sensing information from flowing blood or dynamic speckles by searching for time- and/or space-invariant information within speckles. Furthermore, the broader impact of SURE applies to a class of light scattering problems, such as optical communications \cite{krenn2016twisted}, remote sensing \cite{wang2022far}, and tomography \cite{goy2019high}. We experimentally demonstrate this perspective by performing optical communication within a dynamic multimode fiber (MMF), which exhibits a two-orders-of-magnitude increase in bandwidth. Collectively, SURE offers unique advantages across various disciplines and emerging applications from optical image processing to cryptographic communication. \par

\begin{table}
\renewcommand\arraystretch{1.4}
\caption{Comparison between different speckle processing modalities}\label{tab1}
\begin{tabular}{cccccccc}
\rowcolor{black!10}
& \makecell*[c]{Digital\\filter \\ \cite{michailovich2006despeckling}} & \makecell*[c]{Texture\\analysis\\ \cite{gossage2003texture}} & \makecell*[c]{Speckle\\contrast\\ \cite{boas2010laser}} & \makecell*[c]{Transmission\\matrix\\ \cite{popoff2010measuring,fan2021high,gong2019optical}} & \makecell*[c]{Speckle\\correlation\\ \cite{bertolotti2012non,katz2014non}} & \makecell*[c]{Machine\\learning\\ \cite{li2018deep,rahmani2018multimode,li2018imaging, borhani2018learning}} & \textcolor{magenta}{SURE} \\
\makecell*[c]{Single-shot\\capability} & \checkmark & \checkmark & N/A & \checkmark & \checkmark & \checkmark & \checkmark  \\
\makecell*[c]{Dynamic\\Scattering} & \checkmark & \checkmark & \checkmark & N/A & \checkmark & \checkmark & \checkmark  \\
\makecell*[c]{Phase\\retrieval-free} & \checkmark & \checkmark & \checkmark & N/A & N/A & \checkmark & \checkmark  \\
\makecell*[c]{Multiple \\ scattering} & N/A & N/A & N/A & \checkmark & \checkmark & \checkmark & \checkmark  \\
\makecell*[c]{Complex\\object} & N/A & \checkmark & N/A & \checkmark & N/A & \checkmark & \checkmark  \\
Classification & N/A & \checkmark & N/A & N/A & N/A & \checkmark & \checkmark  \\
Prior-free & N/A & N/A & N/A & \checkmark & N/A & \checkmark & \checkmark  \\
\makecell*[c]{Labeled \\ data size} & None & Several & None & \makecell*[c]{Size\\dependent} & None & Massive & None \\
\hline
\end{tabular}
\end{table}

\section{Results}\label{sec2}
\begin{figure}[htb]
\centering
\includegraphics[width=0.7\textwidth]{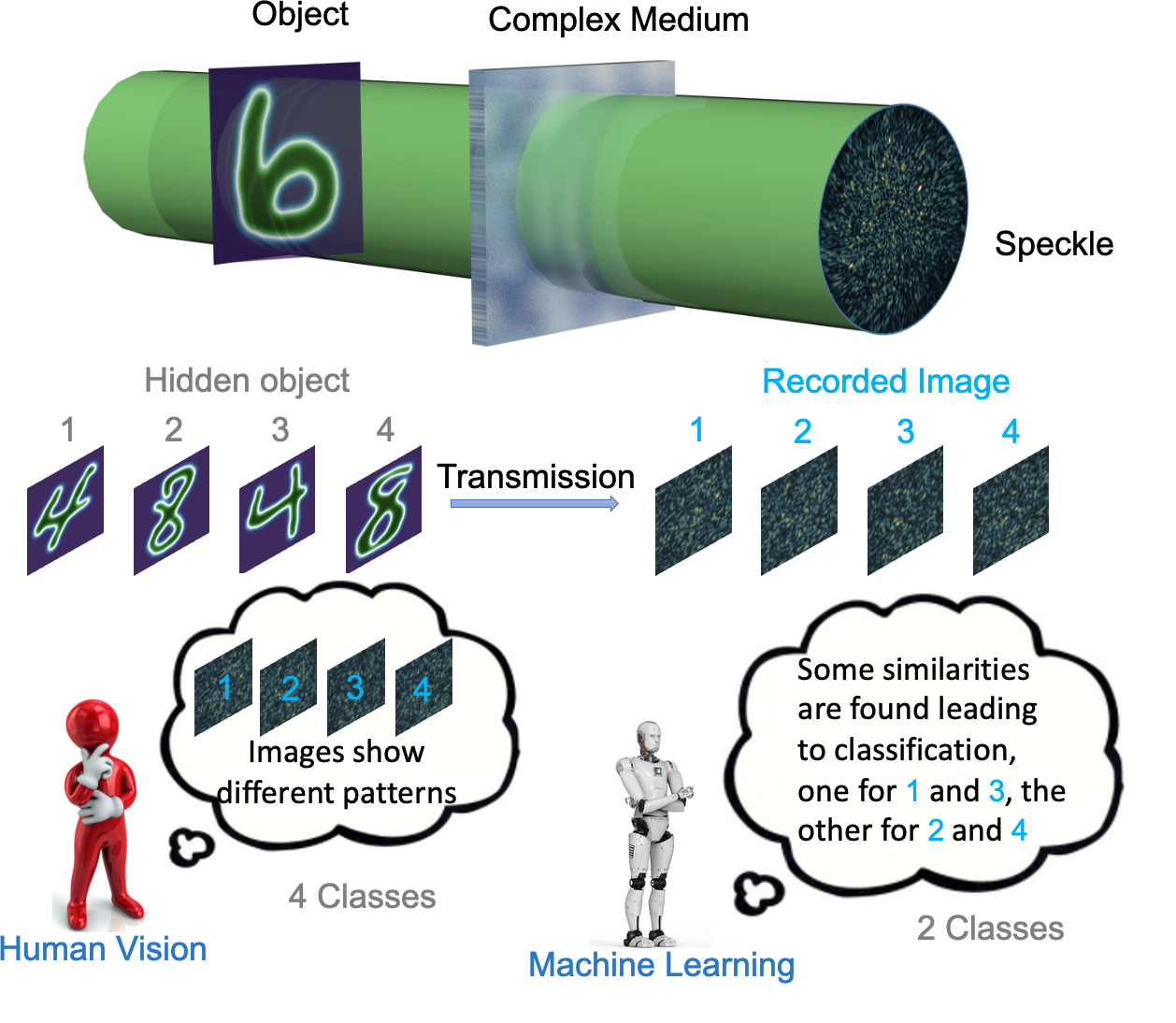}
\caption{The overall concept of speckle clustering for image recognition and the data workflow. Light carrying object information passes through random media yielding corresponding speckles, where a slight perturbation in the wavefront leads to significant speckle decorrelation. These speckles lack salient structural features and exhibit an irregular intensity distribution, such that the original information is scrambled and concealed. Without prior knowledge, it is difficult for human vision to extract the critical information of speckles. While machine learning can accurately extract the intrinsic features of speckles and correctly classify them.}\label{Fig1}
\end{figure}

SURE utilizes the automated clustering within an unsupervised learning framework to reveal internal correlations hidden in these seemingly irrelevant speckles. Remarkably, it extracts information without requiring extensive ground truth or labels that are typically unavailable in speckle interpretation (Fig. \ref{Fig1}). SURE contains two major elements: clustering the speckles and assigning corresponding semantic labels to the resultant clustered groups. The goal of clustering is to discern the high-level correlations within speckles, independent of their specific meanings. The implementation of conventional image clustering is according to the salient structural features from the images, e.g., clustering the handwritten digits, which is also realizable for human vision. However, the input and its corresponding speckles in our case are uncorrelated in spatial structures, such that SURE seeks invariant information to achieve clustering. To achieve this, computational graphs with a self-learning framework are vital to grasp the invariant information facilitating the automated clustering.\par

\begin{figure}[htb]
\centering
\includegraphics[width=\textwidth]{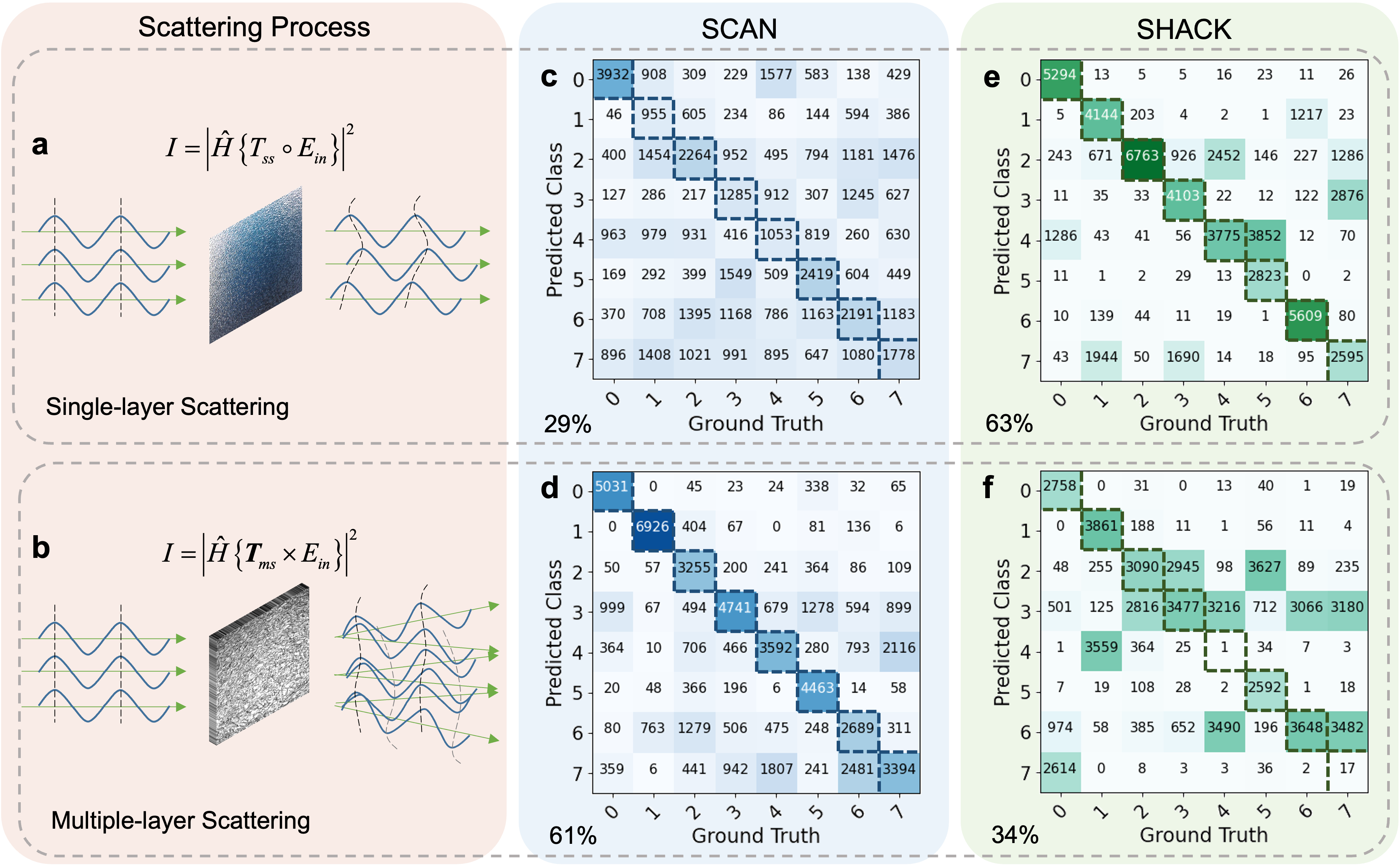}
\caption{Clustering results of different complex media. \textbf{a} and \textbf{b} are wavefront evolutions with single- and multiple-layer scattering, respectively. $\hat{H}$ is the operator describing the propagation of light in free space. \textbf{c} and \textbf{d} are SCAN’s clustering results with speckles generated by single-layer and multiple-layer scattering, respectively. \textbf{e} and \textbf{f} are the same as \textbf{c} and \textbf{d}, but SHACK is employed to perform clustering, respectively. The dashed box are true positives and the number of each subgraph in the lower left corner represents the accuracy. $\circ$ is the Hadamard product, $\mid\cdot\mid$ takes the amplitude of electric field.}\label{Fig2}
\end{figure}

The efficiency of computational graphs for extracting invariant information is discrepant under different physical properties of scattering media \cite{bender2021circumventing}. Therefore, it is crucial to design appropriate computational graphs to implement SURE. Generally, linear scattering media can be categorized into single/thin and multiple/thick scattering layers \cite{rotter2017light}. In single-layer scattering, light passing through the media, e.g., ground glass, water mist and thin biological tissue, merely undergoes a random phase shift at different spatial positions but its direction remains unchanged (Fig. \ref{Fig2}a). Consequently, the original information carried by the beam is still locally embedded in the same positions/pixels before and after the media, defined as local embeddedness. This process can be described by a Hadamard product between input field $E_{in}$ and the matrix $T_{ss}$ with the same number of complex elements. In contrast, with multiple-layer scattering media, e.g., egg membrane, plastered wall and MMF, the change simultaneously appears in both random phase shift and propagation direction (Fig. \ref{Fig2}b). The global information is carried throughout in this case because each speckle grain arises from a coherent superposition of all elements on the incident light, defined as global embeddedness. It can be described by a matrix-vector multiplication between input field $E_{in}$ and complex matrix $\boldsymbol{T}_{ms}$ with $m \times n$, where $m$ is the number of pixels of speckle in the detector and $n$ is the number of elements of input field.\par

We meticulously design computational graphs for each aforementioned category to accommodate SURE and enhance the performance of speckle image sensing. These graphs are termed as speckle clustering aware network (SCAN) and speckle hierarchical agglomeration clustering knack (SHACK), respectively (see Methods and Extended Data Fig. \ref{FigS1} for details). The former is based on a deep neural network which is suitable for processing speckles with global embeddedness, because it has the capability to perceive and capture the high-level correlation among speckles. While the SHACK employs a binary tree to propel the speckle image sensing with local embeddedness. Similar to traditional learning-based methods, SCAN and SHACK convert the speckle clustering into an optimization question to process complex data. The optimization targets are to iteratively minimize the objective function. SCAN’s objective function has the form of $ \mathrm{argmin} \big\{ \sum_i  \mathcal{L} \big( F_{\mathit{NN}}(x_i),F_{\mathit{NN}}(\mathcal{T}(x_i))\big) \big\}$ and SHACK’s is the $ \mathrm{argmin} \big\{ \sum_{ij} d(x_i,x_j) \big\}$, where the $F_{\mathit{NN}}(\cdot)$ represents neural network for SCAN, $\mathcal{T}(\cdot)$ is a random transformation of the original speckle pattern, $\mathcal{L}(\cdot)$ is the loss function for SCAN, $d(x_i,x_j)$ is the distance between $x_i$ and $x_j$, and $\mathrm{argmin} \{ \cdot \}$ is the value of the variable that minimizes the following expression. See Methods for principles and details.\par

To demonstrate the performance of these two frameworks, we implemented the automatic clustering for speckle image sensing from the single- and multiple-layer scattering media (Extended Data Fig. \ref{FigS2} for experimental setup). The quantitative results are shown in Fig. \ref{Fig2}c-f by confusion matrices. These results indicate that SHACK has a higher accuracy for processing single-layer scattering, where the results are dominated by the diagonal terms of the confusion matrix, however, SCAN involves all matrix elements. Complementarily, SCAN effectively handles the speckle generated from the multiple-layer scattering, with its results focusing on the diagonal terms of the confusion matrix, while SHACK outputs a series of random values. These frameworks collectively characterize the performance of speckle interpretation across diverse scenarios by SURE. To further validate the universality of SURE, the clustering is accomplished with different number of classes (tasks with 2, 4 and 8 classes), and the results are consistent with the previous analysis (shown in Extended Data Fig. \ref{FigS3}).\par

It is worth pointing out that SCAN and SHACK can have comparable performance in speckle image sensing when more computing resources are invested to improve the generality of SURE, such that the deliberately selecting either of them is unnecessary. This allows us to bolster the execution of speckle image sensing by expanding the computational graphs. For instance, an improved performance of SCAN in single layer scattering can be achieved by increasing the parameter scale, e.g., the channel number of the convolutional layer and the depth of the network. Additionally, the combination of data dimension reduction methods and SHACK can also be used to enhance the extraction of abstract invariant information from speckles. Furthermore, our results indicate an accuracy surpassing 50\% for categorization tasks with 8 classes, compared to a baseline of 12.5\% for random guessing (see Fig. \ref{Fig2}). This means that this accuracy can be further increased by involving ensemble learning with the boosting techniques \cite{sagi2018ensemble, zhou2021ensemble}.\par

Once the speckle image sensing is achieved by SURE, various applications can be developed by embedding SURE into their corresponding pipelines. As a proof-of-concept demonstration, we propose a SURE-based noninvasive glucose monitoring. In clinical practice, invasive blood collection is still the mainstream method thanks to its high precision, convenience and low cost \cite{bilous2021handbook}. However, a definite diagnostic of diabetes requires a series of testing data, leading to repetitive blood collection. It not only causes secondary injuries to patients, but also increases infection risks. Light-based stand-off methods for glucose detection are promising in biomedical applications because of their compatibility and selectivity \cite{steiner2011optical, tuchin2008handbook}. Amidst the current optical noninvasive blood glucose testing, one of the key challenges is that the skin tissue prevents light penetration.\par

When the probe light passes through the biological tissues \cite{ozana2014improved, ozana2018remote}, the strong scattering leads to signals hiding in the complex speckle background, and hence poses a challenge for glucose monitoring. To address this issue, the combination of ML and optical apparatuses enables the extraction of the glucose signal from speckle backgrounds with a high signal-to-noise ratio \cite{pal2022non, gusev2020noninvasive}. However, a large number of labeled samples is needed to train an ML model for extracting relevant features to achieve anomaly detection and diagnosis due to the intricacy and enormity of glucose monitoring \cite{pal2023noninvasive}. Moreover, on account of the individual dependence on these features, such noninvasive blood glucose monitoring requires multiple invasive blood collections every day to build a training dataset for an individual patient, which causes repeated injuries and infection risks \cite{tang2020non}. Thus, a scattering-based noninvasive glucose monitoring with minimal injuries is urgently in demand.\par

\begin{figure}[htb]
\centering
\includegraphics[width=\textwidth]{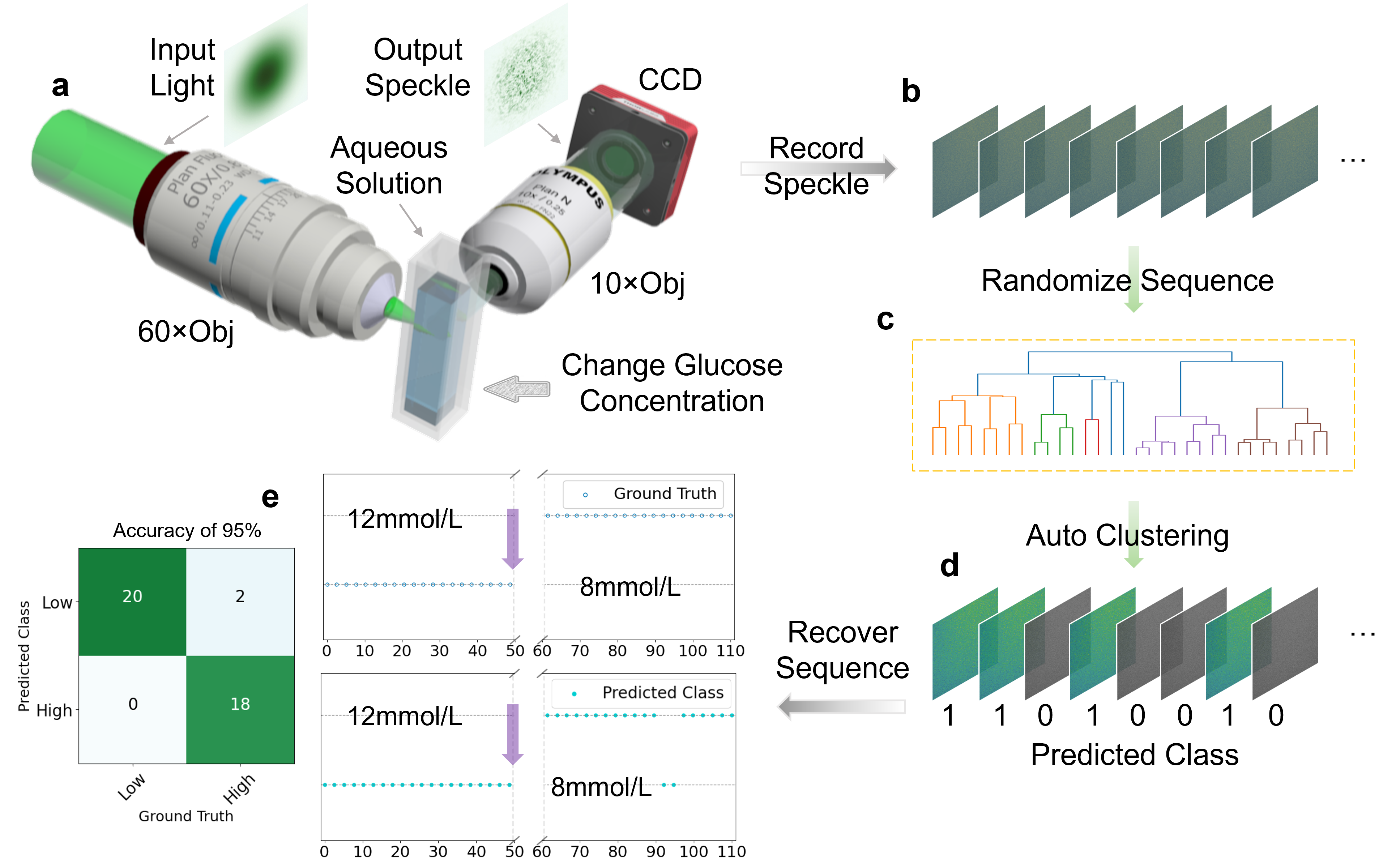}
\caption{Experimental schematics and results of SURE-based high-sensitive noninvasive glucose monitoring. \textbf{a,} Experimental setup. A weak laser beam is focused on the sample, and the speckle patterns are collected at a right angle. \textbf{b,} Random shuffle of the image sequence. \textbf{c,} Calculated graph. \textbf{d,} Results of predicted classes. \textbf{e,} Quantitative results with 20 sets of data for each concentration of sample. The hollow blue dots and the solid blue dots represent the ground truth and the predicted results, respectively. To simulate the scattering of biological tissues, the rear surface of the cuvette is covered with an egg membrane, and the other three surfaces are affixed with Scotch tape (the actual sample is shown in see Extended Data Fig.  \ref{FigS4}).}\label{Fig3}
\end{figure}

Towards this goal, we experimentally validated the SURE-based noninvasive glucose monitoring in a right-angle scattering configuration, in which a biological membrane was introduced to mimic skin tissues (Fig. \ref{Fig3}a). A series of speckle patterns were collected from the aqueous solutions for a fixed glucose concentration. We repeated the same procedure for different glucose concentrations, and these recorded speckle patterns were randomly shuffled in their sequence and fed into the computational graph (SCAN/SHACK) (Fig. \ref{Fig3}b and c). The computational graph automatically sought the high-level features of speckles to achieve the clustering (Fig. \ref{Fig3}d). After that, the true concentrations were assigned to different classes depending on the tagged speckles, where an additional one-time invasive collection was implemented to label one speckle within each class. Note that the speckles exhibit minute random variations even in a concentration-uniform solution, due to local concentration fluctuations of the solution \cite{euliss1984dynamic, yoshizaki2003dielectric, wu1991enhanced}. Such a random effect (like blood flow) leads to time-dependent speckles, complicating concentration classification in the actual scenario. Strikingly, since SURE is based on the invariant information, our strategy was found to be immune to speckle shifts arising from local fluctuations.\par

As a result, we demonstrated 95\% accuracy in identifying the classes of concentration from the time-dependent speckles (Fig. \ref{Fig3}e). Here, two classes of concentration were used for proof-of-concept demonstration, i.e., 8mmol/L and 12mmol/L which are approximately within the blood glucose ranges of healthy people and diabetes patients. In practical applications, more levels of glucose concentration can be involved by adjusting the preset class numbers, e.g., 8 classes (some results see Extended Data Fig. \ref{FigS5}). In addition, one concern in data collection is that the speckle patterns exhibit a time correlation, i.e., speckles that are closer in the time domain tend to have a more similar spatial distribution. Therefore, we need to exclude the possibility that the clustering merely reflects such time correlation. To realize this, we increased the time interval of data acquisition and performed a control experiment in which the glucose solution was replaced with clear water while others remained unchanged. The results of the control group revealed an accuracy of 60\%, which is close to the random accuracy baseline (50\%). Thus, the clustering results can be attributed to different concentrations rather than time correlation (see Extended Data Fig. \ref{FigS4} for additional data and comparisons). It is also worth noting that other components in the body can also be detected and monitored following a similar procedure.\par

\begin{figure}[htb]
\centering
\includegraphics[width=\textwidth]{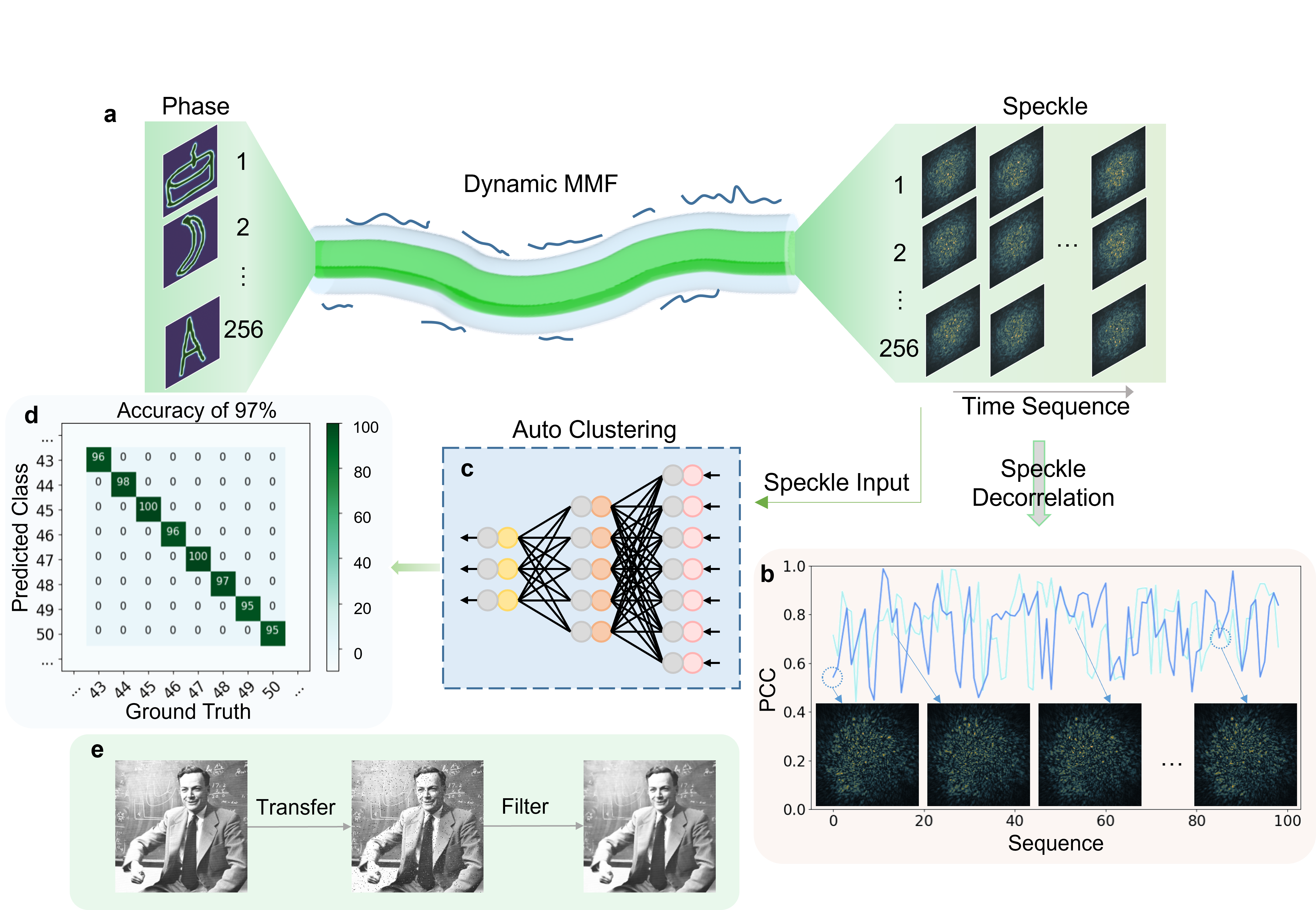}
\caption{SURE-based high-throughput dynamic MMF information transmission. \textbf{a,} For a defined wavefront of light, the destabilization of the MMF results in significantly different speckle patterns. \textbf{b,} Speckle similarity of the same modulation phase at a different time. The light blue and dark blue lines represent the correlation curves of two different phase maps, respectively. \textbf{c,} Calculated graph for clustering time-dependent speckles. \textbf{d,} A confusion matrix of 10 speckle clustering results, randomly intercepted from 256 encoded phase maps. \textbf{e,} The result of the photograph transport. The median filter is employed to remove impulse noise. Additional data are available in Extended Data Fig. \ref{FigS6} and \ref{FigS7}.}\label{Fig4}
\end{figure} 

To further demonstrate the versatility, SURE was deployed into optical communication to achieve high-throughput information transport under dynamic multimode fiber (MMF). Optical fibers serve as the backbone of telecommunication which are ubiquitous in modern society \cite{hecht2004city}. Over the past decades, the data-carrying capacity of fibers has surged by several orders of magnitude through various multiplexing techniques, in which the optical power injected into fibers is growing higher and higher \cite{bozinovic2013terabit}. However, the higher optical power used, the stronger nonlinear effect produced by optical fibers \cite{bozinovic2013terabit,willner2019optical}, severely constraining the continued increase in bandwidth of optical communication. As a solution, we propose a novel scheme to improve the bandwidth, where the wavefront phase is used to encode information while SURE is employed to decode the signal. In this strategy, the different spatial phase distributions can generate significantly different speckle patterns \cite{li2018deep}, such that the mutually orthogonal modulations of input wavefront are unnecessary. Therefore, leveraging the wavefront phase can be conducive to high-bandwidth information transport.\par

In long-distant communication, a confronting challenge arises from multiple sources, e.g., temperature variance and mechanical vibrations, causing instability for MMF and hence resulting in alterations to the output speckles with the identical input (Fig. \ref{Fig4}a and b). To mimic turbulence/disturbance, we imposed random motion on the MMF during the optical communication. In the procedure of transport, the wavefront phases were firstly one-by-one transmitted according to the allocated sequence, and the numerical orders and the corresponding speckles were used as a secret key or protocol to decode speckles and assign values. Thereafter, based on this protocol (the numerical orders of speckles in here), the original data was encoded and transmitted to yield the corresponding speckles. These speckles and those in the protocol were mixed and fed into SCAN/SHACK to perform the clustering and assign values to the clustering group according to the speckles in protocol (Fig. \ref{Fig4}c). This process completed the data transport using the dynamic MMF. The quantitative results are shown by the confusion matrix with the accuracy of 97\% (Fig. \ref{Fig4}d). This result indicates that SURE can effectively uncover the invariant features within dynamic speckles for clustering, enabling high-precision information transport.\par

To further demonstrate the effectiveness of SURE in optical communication, which often involves image transmission, we present an example featuring a photograph of Richard Feynman, with the corresponding results shown in Fig. \ref{Fig4}e. Specifically, a series of phase maps were arbitrarily chosen and numbered in ascending order as the pixel/data values. Similar to the previous strategy, the speckles were acquired at the output end of the MMF and then mixed with the previous speckles in protocol to cluster and assign pixel values. In parallel with current multiplexing techniques, we conducted a random intensity attenuation to each recorded speckle to simulate the coexisting case of intensity and phase encoding. The result of the image transport exhibited high fidelity except for several anomalous pixels. These anomalies are characterized by impulse noise which is tractable by the median filter (Fig. \ref{Fig4}e). This indicates that the SURE-based technique can effectively implement MMF-based optical communication and perform speckle sensing under dynamic scattering. Moreover, this scheme can also be extended to medical endoscopic widefield imaging with a single MMF.\par

\section{Discussion}\label{sec3}
The SURE methodology provides a conceptual framework for speckle image sensing by extracting abstract, invariant features from complex and noisy speckle backgrounds. Taking advantage of unsupervised learning, SURE, as demonstrated in numerous examples shown above, eliminates the extensive labeled dataset for training, and effectively handles the speckle classification under static and dynamic environments regardless of the amount of scattering. As demonstrated in two applications, SURE correctly classifies speckles associated with different glucose concentrations and distinguishes images being transmitted through MMF, thus achieving noninvasive glucose monitoring and enabling a high-throughput optical communication. Compared with previous methods, SURE-based glucose monitoring discards the cumbersome measuring apparatus and intricate algorithm, substantially simplifying the diagnosis procedure and holding promise for miniaturized and portable noninvasive glucose monitors. In optical communications, SURE is proven to sort the speckle data under dynamic scattering, significantly improving the noise resilience and throughput of MMF-based information transport. It is important to emphasize that SURE is not confined to specific applications, but rather serves as a universal solution for diverse challenges where specific information extraction from complex patterns is necessary. For example, it can be used for defect detection on chips, encryption and decryption of optical information, and allow signal and image recognition with single/low photon illumination, i.e., it can be potentially extended to quantum imaging and biological imaging, where optical damage to fragile biological samples needs to be avoided.\par

To implement SURE, SCAN/SHACK employ a contrastive learning strategy to capture the context of speckle. This usually requires data with alignment and uniformity to ensure grasping salient features among similar instances and discriminating the different instances \cite{wang2020understanding}. These can be boosted by imposing physical constraints and preprocessing speckle patterns. The former can be satisfied by only setting one factor needed for observation to change speckle while ensuring the others remain unchanged or uniformly varied. The latter is achieved by introducing external physical priors or resorting to the state-of-the-art pretrained models. Furthermore, designing the positive and negative samples is a practical approach to facilitate SURE \cite{momeni2023backpropagation}. It is also worth discussing SURE’s two steps of clustering and label assignment. In the first step, although the clustering method is tailored for tasks involving discrete objects, auto-encoding models can be introduced to handle tasks with continuous variables. In the second step, SURE needs to integrate with existing methods or well-designed protocols to assign meaningful labels to cluster groups for specific downstream tasks. Moreover, we point out that SURE needs to select an appropriate framework (SCAN/SHACK) according to specific tasks, due to the limitations of computational resources. With ongoing advancements in both software and hardware, SURE is expected to adeptly address image sensing and speckle interpretation challenges generating new far-reaching applications.\par

\section{Conclusion}\label{sec4}
In conclusion, we proposed, designed, experimentally demonstrated, and validated a new paradigm in speckle sensing, namely speckle unsupervised recognition and evaluation (SURE), which is based on acquiring speckle patterns and utilizing an unsupervised learning algorithm to reveal invisible information from speckle patterns. Firstly, we prove that SURE can faithfully cluster the speckle corresponding to the same class of input field, regardless of whether there is single-layer scattering or multiple-layer scattering. After that, incorporating SURE into the pipeline of speckle interpretation tasks, we demonstrate various crucial applications in speckle image sensing in disparate disciplines, i.e., noninvasive glucose monitoring and harsh-condition MMF-based communication. The establishment of SURE provides a new route to comprehend optical speckles and disordered systems and is promising to benefit numerous tasks within modern photonics, such as deep tissue imaging, automatic driving and remote sensing, across various scientific and technological fields.\par


\section*{Methods}\label{sec5}
\subsection*{Experimental Details}
\textit{Experimental Setup}: As shown in Extended Data Fig. \ref{FigS2}, a continuous wave with a 532~nm wavelength was generated by laser (YFA-SF-1064-50-CW, Precilaser), and injected into a single-mode optical fiber with a length of 3m to shape the beam profile. To match the polarization state of the spatial light modulator (SLM; X13139-09, Hamamatsu), a zero-order half-wave plate (HWP) and a polarizing beam splitter (PBS) were inserted to produce the polarization along the SLM and control the power of the light. In the case of scattering media (ZnO with 400~$\mu$m thickness or ground glass with 600 grit), the beam was focused by an objective lens (20$\times$/0.4NA, Olympus). For MMF (SR-opt-8039, 2~m length, $\varnothing$105~$\mu$m, 0.22NA, Andor) and glucose aqueous solution, the beam was focused through an objective lens with a larger numerical aperture (50$\times$/0.6NA, Nikon). Then, speckles were collected by an objective lens (10$\times$/0.3NA, Olympus) and recorded by a charge-coupled device (CCD; Prosilica GT1910, AVT) for further subsequent information extraction.\par

\textit{Dynamic environment for MMF}: A MMF with a length of 2~m was wound into coils with a 10~cm diameter. To mimic environmental disturbances, an inhomogeneous stress was imposed on such coils by a motorized rotator with a constant velocity of 20~degrees/s and two loose clamps, where friction was randomly applied to the MMF. The collected speckles were randomly attenuated in time to verify the compatibility of SURE with intensity modulation for optical communications. When the nonlinear effect of the fiber can be ignored, the time-sequence random attenuation of the collected data is equivalent to directly attenuating the power of light.\par

\textit{Glucose solution preparation}: The glucose stock solution was prepared with a concentration of 0.305~mol/L (D-glucose: 549~mg, deionized water: 10~mL). In the experiment, 3~mL of water and 0.08~mL of the stock solution were added into the cuvette to obtain an 8~mmol/L glucose aqueous solution corresponding to the average level of a healthy people. Subsequently, 0.04~mL of the stock solution was continued to be added to obtain a 12~mmol/L corresponding to the average level of patient with hyperglycemia. Noted that the cuvette was wrapped with two layers of scotch tape to simulate thin scattering media. The rear surface of the cuvette had an extra layer of egg membrane as a thick backscattering layer.\par

\subsection*{Data acquisition}
In the task of speckle image clustering, the MNIST dataset was used as the modulation phase map, where the size of each image was interpolated from 28$\times$28 to 400$\times$400 using bicubic interpolation. The recorded speckles had 256×256 pixels with 8-bit grayscale. All MNIST data was loaded into the SLM one by one, and 70,000 speckle images were acquired to build the dataset. For the experiment of information transport based on dynamic MMF, the DIV2K dataset was adopted to modulate the phase with a total of 800 images, each appearing 100 times. The size of each phase map was 256$\times$256 pixels, and the corresponding speckles had 256$\times$256 pixels (8-bit). Here, 256 classes were randomly selected to build a dataset with 25,600 speckles. For noninvasive glucose monitoring, the acquisition interval of each image was 5 seconds with 1024$\times$1024 pixels (16-bit). Since the change in speckle arising from the change in glucose concentration was slight compared with the influence of the environmental disturbance, we grouped 25 continuous frames of speckles to construct one video, which can facilitate SURE for information extraction from speckles.\par

\subsection*{SCAN implementation}
\textit{Structure}: SCAN utilizes invariant information to maximize mutual information for clustering. Its basic structure comprises a convolutional neural network (CNN) for feature extraction, a fully connected (FC) layer for clustering and an overclustering module to enhance feature extraction. In the clustering procedure, speckles are cropped at the center to single-channel 64$\times$64 pixels and perform random transformations. The pre-processing speckle images are then fed into the CNN consisting of 3$\times$3 convolution, Batch Normalization, Leaky ReLU and average pooling, to extract feature maps with 1024 channels and 8$\times$8 pixels. Subsequently, the feature maps are flattened and fed into the FC layer (clustering or overclustering module) to yield the probability for each batch over the relevant clusters. The input neurons of the FC layer are 1024$\times$8$\times$8 and the output neurons are matched to the number of classes. To enhance the stability of the training process, multiple clustering modules with disparate initialization processes are employed to collaboratively fine-tune the parameters of SCAN.\par

\textit{Training}: The PyTorch framework was used to build the SCAN model, and its training and evaluation were implemented on a server (Intel Xeon Gold 6248 CPU, Tesla V100 PCIe 32GB GPU, CUDA10.2). Before training, the auxiliary dataset is created by random transformations of the original dataset, such as scaling, skewing, rotating and flipping, to facilitate the CNN in capturing the invariant information from speckles. During training, the CNN has two inputs that are sampled from the original and auxiliary datasets, and are separately fed into the CNN to generate the corresponding feature maps for the subsequent FC layer. When the original and transformed data are grouped into the same class, it indicates that the CNN has extracted crucial invariant features from speckles and the FC layers can accurately classify them. To further improve the effectiveness of feature extraction, SCAN incorporates an overclustering module which is also an FC structure but has a significantly larger number of output classes than the preset cluster number. The aim of the overclustering module is still to group the original and transformed data into the same class. Since more classes are introduced, the classification requires more precise and sufficient features to be extracted by the CNN. The training process alternately utilizes the overclustering and clustering modules in each epoch. The best clustering module (lowest loss) is selected to serve as the model output.\par
In SCAN, we select the IIC-Loss function as loss function, which is defined as \cite{ji2019invariant},
\begin{equation}
\mathcal{L} = -\sum_{jh} \Big\{ P_{jh} \cdot \big[\ln P_{jh} - \ln (P_j P_h)  \big] \Big\}
\label{equ1}
\end{equation}
\begin{equation}
P_{jh} = \frac{1}{n} \sum_{i=1}^n \big [ P_i(y = j) \cdot P_i(y^{\prime} = j) \big]
\label{equ2}
\end{equation}
where $P_i \in [0, 1]^C$ is the probability of network output corresponding to each cluster of the $i$-th sample pairs in each batch (batch size is $n$) and $C$ is the total number of clusters. Since there are two ways to sample a data pair, $(x, x^{\prime})$ or $(x^{\prime}, x)$, we only considered the symmetry problem with $\boldsymbol{P} = (\boldsymbol{P} + \boldsymbol{P}^T) / 2$ where $\boldsymbol{P}$ is a matrix with elements of $P_{jh}$. The marginal probabilities $P_j = P(y = j)$ and $P_h = P(y^{\prime} = h)$ can be calculated by summing the rows and columns of $\boldsymbol{P}$. The CNN module, clustering module, and over-clustering module are weight-sharing in each stage. To improve the robustness of SCAN, we parallelly trained multiple clustering modules and the loss of all clustering modules was averaged to serve as the final loss. The Adam optimizer was run for 40 epochs with the initial learning rate of $10^{-4}$.\par

\textit{Evaluation}: Due to the absence of the ground truth, when one speckle image is input, the neural network can merely predict the corresponding probability of each clustering group without the actual meanings of the class. This leads to the results of SCAN being disorderly distributed in the confusion matrix, such that it is difficult to determine the correspondence between SCAN's output numbers and the actual label numbers. To address this, SCAN utilized linear assignment to find the best one-to-one permutation mapping of the predictions of the testing dataset. Here, we let $\boldsymbol{M}$ be a $C \times C$ matrix with columns representing the predicted clusters and rows representing the true clusters of the sample. In the linear assignment method, $\boldsymbol{X}$ is assumed to be a Boolean matrix of $C \times C$. If the $j$-th predicted label corresponds to the $h$-th true label, then the value of $X_{jh}$ is 1. Once the optimal mapping is achieved, the corresponding $\boldsymbol{X}$ satisfies the maximization of $\sum_{jh} M_{jh} X_{jh}$. The “linear\_sum\_assignment” function in the SciPy library was employed to complete the linear assignment and return the best mapping. To quantitatively evaluate the clustering performance, the ground truth was introduced in this step without being presented in the training process. The accuracy of clustering was defined as the ratio of true positives to total samples, which is used to evaluate the performance of SURE.\par

\subsection*{SHACK implementation}
SHACK is an agglomeration clustering algorithm that creates a hierarchical nested clustering tree by calculating the similarity among all speckles. Each sample is treated as a unit in SHACK, and two units with the highest similarity are grouped together. As a result, the total number of units decreases by one. This process is repeated until the number of clusters matches the ground truth number of clusters. Concretely, for each step of hierarchical clustering, speckle images are paired in all possible combinations. The paired images are treated as a new unit and the variance for all pixel values within the unit is calculated (initially, the variance of each sample is set to 0). The difference in variance between after and before pairing is defined as the cost function. Traverse all possible pairs and select the pair with the smallest cost function as the result of the current clustering step. The "AgglomerativeClustering" class of Scikit-Learn library was used to build SHACK with the Euclidean distance for measuring the similarity. The speckle images were down-sampled to 64$\times$64 pixels, then flattened and fed into SHACK for training. Similarly, the linear assignment was employed to find the best one-to-one permutation mapping.

\backmatter




\section*{Declarations}

\bmhead{Acknowledgements}
This work was supported by the National Natural Science Foundation of China (Grant No. 11934011, 12074339, 62075194, U21A6006, 62202418, U21B2004), the National Key Research and Development Program of China (Grant No. 2019YFA0308100, 2023YFB2806000, 2022YFA1204700), the Strategic Priority Research Program of Chinese Academy of Sciences (Grant No. XDB28000000), the Leading Innovation and Entrepreneurship Team in Zhejiang Province (Grant No. 2020R01001), the Open Program of the State Key Laboratory of Advanced Optical Communication Systems and Networks at Shanghai Jiao Tong University (Grant No. 2023GZKF024), the Fundamental Research Funds for the Central Universities, the Information Technology Center and State Key Lab of CAD\&CG, the Zhejiang Provincial Key Laboratory of Information Processing, Communication and Networking (IPCAN), the Air Force Office of Scientific Research (AFOSR) (FA9550-20-1-0366, FA9550-23-1-0599), and the National Institutes of Health (NIH) (R01GM127696, R01GM152633, R21GM142107, and 1R21CA269099).

\bmhead{Competing interests}
The authors declare no competing interests.

\bmhead{Data availability}
All data used to produce the findings of this study are available from the corresponding author on reasonable request.

\bmhead{Code availability}
Codes for the whole pipeline of this study are available via GitHub at \url{https://github.com/weirufan/SURE}.

\bmhead{Author contribution}
W.F., D.W.W. and D.Z. conceived the idea and designed the experiment. W.F. carried out the experiment, collected data and performed the theoretical modeling of SURE. X.T. wrote and implemented the SCAN and SHACK. W.F., X.T., V.V.Y., D.W.W. and D.Z. analyzed data and wrote the manuscript. The project is supervised under D.W.W. and D.Z. All authors discussed the results and revised the manuscript.



\begin{thebibliography}{55}
\ifx \bisbn   \undefined \def \bisbn  #1{ISBN #1}\fi
\ifx \binits  \undefined \def \binits#1{#1}\fi
\ifx \bauthor  \undefined \def \bauthor#1{#1}\fi
\ifx \batitle  \undefined \def \batitle#1{#1}\fi
\ifx \bjtitle  \undefined \def \bjtitle#1{#1}\fi
\ifx \bvolume  \undefined \def \bvolume#1{\textbf{#1}}\fi
\ifx \byear  \undefined \def \byear#1{#1}\fi
\ifx \bissue  \undefined \def \bissue#1{#1}\fi
\ifx \bfpage  \undefined \def \bfpage#1{#1}\fi
\ifx \blpage  \undefined \def \blpage #1{#1}\fi
\ifx \burl  \undefined \def \burl#1{\textsf{#1}}\fi
\ifx \doiurl  \undefined \def \doiurl#1{\url{https://doi.org/#1}}\fi
\ifx \betal  \undefined \def \betal{\textit{et al.}}\fi
\ifx \binstitute  \undefined \def \binstitute#1{#1}\fi
\ifx \binstitutionaled  \undefined \def \binstitutionaled#1{#1}\fi
\ifx \bctitle  \undefined \def \bctitle#1{#1}\fi
\ifx \beditor  \undefined \def \beditor#1{#1}\fi
\ifx \bpublisher  \undefined \def \bpublisher#1{#1}\fi
\ifx \bbtitle  \undefined \def \bbtitle#1{#1}\fi
\ifx \bedition  \undefined \def \bedition#1{#1}\fi
\ifx \bseriesno  \undefined \def \bseriesno#1{#1}\fi
\ifx \blocation  \undefined \def \blocation#1{#1}\fi
\ifx \bsertitle  \undefined \def \bsertitle#1{#1}\fi
\ifx \bsnm \undefined \def \bsnm#1{#1}\fi
\ifx \bsuffix \undefined \def \bsuffix#1{#1}\fi
\ifx \bparticle \undefined \def \bparticle#1{#1}\fi
\ifx \barticle \undefined \def \barticle#1{#1}\fi
\bibcommenthead
\ifx \bconfdate \undefined \def \bconfdate #1{#1}\fi
\ifx \botherref \undefined \def \botherref #1{#1}\fi
\ifx \url \undefined \def \url#1{\textsf{#1}}\fi
\ifx \bchapter \undefined \def \bchapter#1{#1}\fi
\ifx \bbook \undefined \def \bbook#1{#1}\fi
\ifx \bcomment \undefined \def \bcomment#1{#1}\fi
\ifx \oauthor \undefined \def \oauthor#1{#1}\fi
\ifx \citeauthoryear \undefined \def \citeauthoryear#1{#1}\fi
\ifx \endbibitem  \undefined \def \endbibitem {}\fi
\ifx \bconflocation  \undefined \def \bconflocation#1{#1}\fi
\ifx \arxivurl  \undefined \def \arxivurl#1{\textsf{#1}}\fi
\csname PreBibitemsHook\endcsname

\bibitem[\protect\citeauthoryear{Rotter and Gigan}{2017}]{rotter2017light}
\begin{barticle}
\bauthor{\bsnm{Rotter}, \binits{S.}},
\bauthor{\bsnm{Gigan}, \binits{S.}}:
\batitle{Light fields in complex media: Mesoscopic scattering meets wave control}.
\bjtitle{Reviews of Modern Physics}
\bvolume{89}(\bissue{1}),
\bfpage{015005}
(\byear{2017})
\end{barticle}
\endbibitem

\bibitem[\protect\citeauthoryear{Savo et~al.}{2017}]{savo2017observation}
\begin{barticle}
\bauthor{\bsnm{Savo}, \binits{R.}},
\bauthor{\bsnm{Pierrat}, \binits{R.}},
\bauthor{\bsnm{Najar}, \binits{U.}},
\bauthor{\bsnm{Carminati}, \binits{R.}},
\bauthor{\bsnm{Rotter}, \binits{S.}},
\bauthor{\bsnm{Gigan}, \binits{S.}}:
\batitle{Observation of mean path length invariance in light-scattering media}.
\bjtitle{Science}
\bvolume{358}(\bissue{6364}),
\bfpage{765}--\blpage{768}
(\byear{2017})
\end{barticle}
\endbibitem

\bibitem[\protect\citeauthoryear{Redding et~al.}{2012}]{redding2012speckle}
\begin{barticle}
\bauthor{\bsnm{Redding}, \binits{B.}},
\bauthor{\bsnm{Choma}, \binits{M.A.}},
\bauthor{\bsnm{Cao}, \binits{H.}}:
\batitle{Speckle-free laser imaging using random laser illumination}.
\bjtitle{Nature Photonics}
\bvolume{6}(\bissue{6}),
\bfpage{355}--\blpage{359}
(\byear{2012})
\end{barticle}
\endbibitem

\bibitem[\protect\citeauthoryear{Bianco et~al.}{2018}]{bianco2018strategies}
\begin{barticle}
\bauthor{\bsnm{Bianco}, \binits{V.}},
\bauthor{\bsnm{Memmolo}, \binits{P.}},
\bauthor{\bsnm{Leo}, \binits{M.}},
\bauthor{\bsnm{Montresor}, \binits{S.}},
\bauthor{\bsnm{Distante}, \binits{C.}},
\bauthor{\bsnm{Paturzo}, \binits{M.}},
\bauthor{\bsnm{Picart}, \binits{P.}},
\bauthor{\bsnm{Javidi}, \binits{B.}},
\bauthor{\bsnm{Ferraro}, \binits{P.}}:
\batitle{Strategies for reducing speckle noise in digital holography}.
\bjtitle{Light: Science \& Applications}
\bvolume{7}(\bissue{1}),
\bfpage{48}
(\byear{2018})
\end{barticle}
\endbibitem

\bibitem[\protect\citeauthoryear{Michailovich and Tannenbaum}{2006}]{michailovich2006despeckling}
\begin{barticle}
\bauthor{\bsnm{Michailovich}, \binits{O.V.}},
\bauthor{\bsnm{Tannenbaum}, \binits{A.}}:
\batitle{Despeckling of medical ultrasound images}.
\bjtitle{IEEE Transactions on Ultrasonics, Ferroelectrics, and Frequency Control}
\bvolume{53}(\bissue{1}),
\bfpage{64}--\blpage{78}
(\byear{2006})
\end{barticle}
\endbibitem

\bibitem[\protect\citeauthoryear{Mudry et~al.}{2012}]{mudry2012structured}
\begin{barticle}
\bauthor{\bsnm{Mudry}, \binits{E.}},
\bauthor{\bsnm{Belkebir}, \binits{K.}},
\bauthor{\bsnm{Girard}, \binits{J.}},
\bauthor{\bsnm{Savatier}, \binits{J.}},
\bauthor{\bsnm{Le~Moal}, \binits{E.}},
\bauthor{\bsnm{Nicoletti}, \binits{C.}},
\bauthor{\bsnm{Allain}, \binits{M.}},
\bauthor{\bsnm{Sentenac}, \binits{A.}}:
\batitle{Structured illumination microscopy using unknown speckle patterns}.
\bjtitle{Nature Photonics}
\bvolume{6}(\bissue{5}),
\bfpage{312}--\blpage{315}
(\byear{2012})
\end{barticle}
\endbibitem

\bibitem[\protect\citeauthoryear{Choi et~al.}{2022}]{choi2022wide}
\begin{barticle}
\bauthor{\bsnm{Choi}, \binits{Y.}},
\bauthor{\bsnm{Kim}, \binits{M.}},
\bauthor{\bsnm{Park}, \binits{C.}},
\bauthor{\bsnm{Park}, \binits{J.}},
\bauthor{\bsnm{Park}, \binits{Y.}},
\bauthor{\bsnm{Cho}, \binits{Y.-H.}}:
\batitle{Wide-field super-resolution optical fluctuation imaging through dynamic near-field speckle illumination}.
\bjtitle{Nano Letters}
\bvolume{22}(\bissue{6}),
\bfpage{2194}--\blpage{2201}
(\byear{2022})
\end{barticle}
\endbibitem

\bibitem[\protect\citeauthoryear{Gossage et~al.}{2003}]{gossage2003texture}
\begin{barticle}
\bauthor{\bsnm{Gossage}, \binits{K.W.}},
\bauthor{\bsnm{Tkaczyk}, \binits{T.S.}},
\bauthor{\bsnm{Rodriguez}, \binits{J.J.}},
\bauthor{\bsnm{Barton}, \binits{J.K.}}:
\batitle{Texture analysis of optical coherence tomography images: feasibility for tissue classification}.
\bjtitle{Journal of Biomedical Optics}
\bvolume{8}(\bissue{3}),
\bfpage{570}--\blpage{575}
(\byear{2003})
\end{barticle}
\endbibitem

\bibitem[\protect\citeauthoryear{Redding et~al.}{2013}]{redding2013compact}
\begin{barticle}
\bauthor{\bsnm{Redding}, \binits{B.}},
\bauthor{\bsnm{Liew}, \binits{S.F.}},
\bauthor{\bsnm{Sarma}, \binits{R.}},
\bauthor{\bsnm{Cao}, \binits{H.}}:
\batitle{Compact spectrometer based on a disordered photonic chip}.
\bjtitle{Nature Photonics}
\bvolume{7}(\bissue{9}),
\bfpage{746}--\blpage{751}
(\byear{2013})
\end{barticle}
\endbibitem

\bibitem[\protect\citeauthoryear{Antipa et~al.}{2018}]{antipa2017diffusercam}
\begin{barticle}
\bauthor{\bsnm{Antipa}, \binits{N.}},
\bauthor{\bsnm{Kuo}, \binits{G.}},
\bauthor{\bsnm{Heckel}, \binits{R.}},
\bauthor{\bsnm{Mildenhall}, \binits{B.}},
\bauthor{\bsnm{Bostan}, \binits{E.}},
\bauthor{\bsnm{Ng}, \binits{R.}},
\bauthor{\bsnm{Waller}, \binits{L.}}:
\batitle{Diffusercam: lensless single-exposure 3d imaging}.
\bjtitle{Optica}
\bvolume{5}(\bissue{1}),
\bfpage{1}--\blpage{9}
(\byear{2018})
\end{barticle}
\endbibitem

\bibitem[\protect\citeauthoryear{Bertolotti et~al.}{2012}]{bertolotti2012non}
\begin{barticle}
\bauthor{\bsnm{Bertolotti}, \binits{J.}},
\bauthor{\bsnm{Van~Putten}, \binits{E.G.}},
\bauthor{\bsnm{Blum}, \binits{C.}},
\bauthor{\bsnm{Lagendijk}, \binits{A.}},
\bauthor{\bsnm{Vos}, \binits{W.L.}},
\bauthor{\bsnm{Mosk}, \binits{A.P.}}:
\batitle{Non-invasive imaging through opaque scattering layers}.
\bjtitle{Nature}
\bvolume{491}(\bissue{7423}),
\bfpage{232}--\blpage{234}
(\byear{2012})
\end{barticle}
\endbibitem

\bibitem[\protect\citeauthoryear{Katz et~al.}{2014}]{katz2014non}
\begin{barticle}
\bauthor{\bsnm{Katz}, \binits{O.}},
\bauthor{\bsnm{Heidmann}, \binits{P.}},
\bauthor{\bsnm{Fink}, \binits{M.}},
\bauthor{\bsnm{Gigan}, \binits{S.}}:
\batitle{Non-invasive single-shot imaging through scattering layers and around corners via speckle correlations}.
\bjtitle{Nature Photonics}
\bvolume{8}(\bissue{10}),
\bfpage{784}--\blpage{790}
(\byear{2014})
\end{barticle}
\endbibitem

\bibitem[\protect\citeauthoryear{Popoff et~al.}{2010}]{popoff2010measuring}
\begin{barticle}
\bauthor{\bsnm{Popoff}, \binits{S.M.}},
\bauthor{\bsnm{Lerosey}, \binits{G.}},
\bauthor{\bsnm{Carminati}, \binits{R.}},
\bauthor{\bsnm{Fink}, \binits{M.}},
\bauthor{\bsnm{Boccara}, \binits{A.C.}},
\bauthor{\bsnm{Gigan}, \binits{S.}}:
\batitle{Measuring the transmission matrix in optics: An approach to the study and control of light propagation in disordered media}.
\bjtitle{Physical Review Letters}
\bvolume{104}(\bissue{10}),
\bfpage{100601}
(\byear{2010})
\end{barticle}
\endbibitem

\bibitem[\protect\citeauthoryear{Fan et~al.}{2021}]{fan2021high}
\begin{barticle}
\bauthor{\bsnm{Fan}, \binits{W.}},
\bauthor{\bsnm{Chen}, \binits{Z.}},
\bauthor{\bsnm{Yakovlev}, \binits{V.V.}},
\bauthor{\bsnm{Pu}, \binits{J.}}:
\batitle{High-fidelity image reconstruction through multimode fiber via polarization-enhanced parametric speckle imaging}.
\bjtitle{Laser \& Photonics Reviews}
\bvolume{15}(\bissue{5}),
\bfpage{2000376}
(\byear{2021})
\end{barticle}
\endbibitem

\bibitem[\protect\citeauthoryear{Silva et~al.}{2022}]{silva2022signal}
\begin{barticle}
\bauthor{\bsnm{Silva}, \binits{V.B.}},
\bauthor{\bsnm{Andrade De~Jesus}, \binits{D.}},
\bauthor{\bsnm{Klein}, \binits{S.}},
\bauthor{\bsnm{Walsum}, \binits{T.}},
\bauthor{\bsnm{Cardoso}, \binits{J.}},
\bauthor{\bsnm{Brea}, \binits{L.S.}},
\bauthor{\bsnm{Vaz}, \binits{P.G.}}:
\batitle{Signal-carrying speckle in optical coherence tomography: a methodological review on biomedical applications}.
\bjtitle{Journal of Biomedical Optics}
\bvolume{27}(\bissue{3}),
\bfpage{030901}
(\byear{2022})
\end{barticle}
\endbibitem

\bibitem[\protect\citeauthoryear{Li et~al.}{2018}]{li2018deep}
\begin{barticle}
\bauthor{\bsnm{Li}, \binits{Y.}},
\bauthor{\bsnm{Xue}, \binits{Y.}},
\bauthor{\bsnm{Tian}, \binits{L.}}:
\batitle{Deep speckle correlation: a deep learning approach toward scalable imaging through scattering media}.
\bjtitle{Optica}
\bvolume{5}(\bissue{10}),
\bfpage{1181}--\blpage{1190}
(\byear{2018})
\end{barticle}
\endbibitem

\bibitem[\protect\citeauthoryear{Rahmani et~al.}{2018}]{rahmani2018multimode}
\begin{barticle}
\bauthor{\bsnm{Rahmani}, \binits{B.}},
\bauthor{\bsnm{Loterie}, \binits{D.}},
\bauthor{\bsnm{Konstantinou}, \binits{G.}},
\bauthor{\bsnm{Psaltis}, \binits{D.}},
\bauthor{\bsnm{Moser}, \binits{C.}}:
\batitle{Multimode optical fiber transmission with a deep learning network}.
\bjtitle{Light: Science \& Applications}
\bvolume{7}(\bissue{1}),
\bfpage{69}
(\byear{2018})
\end{barticle}
\endbibitem

\bibitem[\protect\citeauthoryear{Li et~al.}{2018}]{li2018imaging}
\begin{barticle}
\bauthor{\bsnm{Li}, \binits{S.}},
\bauthor{\bsnm{Deng}, \binits{M.}},
\bauthor{\bsnm{Lee}, \binits{J.}},
\bauthor{\bsnm{Sinha}, \binits{A.}},
\bauthor{\bsnm{Barbastathis}, \binits{G.}}:
\batitle{Imaging through glass diffusers using densely connected convolutional networks}.
\bjtitle{Optica}
\bvolume{5}(\bissue{7}),
\bfpage{803}--\blpage{813}
(\byear{2018})
\end{barticle}
\endbibitem

\bibitem[\protect\citeauthoryear{Borhani et~al.}{2018}]{borhani2018learning}
\begin{barticle}
\bauthor{\bsnm{Borhani}, \binits{N.}},
\bauthor{\bsnm{Kakkava}, \binits{E.}},
\bauthor{\bsnm{Moser}, \binits{C.}},
\bauthor{\bsnm{Psaltis}, \binits{D.}}:
\batitle{Learning to see through multimode fibers}.
\bjtitle{Optica}
\bvolume{5}(\bissue{8}),
\bfpage{960}--\blpage{966}
(\byear{2018})
\end{barticle}
\endbibitem

\bibitem[\protect\citeauthoryear{Wang et~al.}{2023}]{wang2023image}
\begin{barticle}
\bauthor{\bsnm{Wang}, \binits{T.}},
\bauthor{\bsnm{Sohoni}, \binits{M.M.}},
\bauthor{\bsnm{Wright}, \binits{L.G.}},
\bauthor{\bsnm{Stein}, \binits{M.M.}},
\bauthor{\bsnm{Ma}, \binits{S.-Y.}},
\bauthor{\bsnm{Onodera}, \binits{T.}},
\bauthor{\bsnm{Anderson}, \binits{M.G.}},
\bauthor{\bsnm{McMahon}, \binits{P.L.}}:
\batitle{Image sensing with multilayer nonlinear optical neural networks}.
\bjtitle{Nature Photonics}
\bvolume{17}(\bissue{5}),
\bfpage{408}--\blpage{415}
(\byear{2023})
\end{barticle}
\endbibitem

\bibitem[\protect\citeauthoryear{Boas and Dunn}{2010}]{boas2010laser}
\begin{barticle}
\bauthor{\bsnm{Boas}, \binits{D.A.}},
\bauthor{\bsnm{Dunn}, \binits{A.K.}}:
\batitle{Laser speckle contrast imaging in biomedical optics}.
\bjtitle{Journal of Biomedical Optics}
\bvolume{15}(\bissue{1}),
\bfpage{011109}--\blpage{011109}
(\byear{2010})
\end{barticle}
\endbibitem

\bibitem[\protect\citeauthoryear{Wang et~al.}{2021}]{wang2021deep}
\begin{barticle}
\bauthor{\bsnm{Wang}, \binits{Y.}},
\bauthor{\bsnm{Louie}, \binits{D.C.}},
\bauthor{\bsnm{Cai}, \binits{J.}},
\bauthor{\bsnm{Tchvialeva}, \binits{L.}},
\bauthor{\bsnm{Lui}, \binits{H.}},
\bauthor{\bsnm{Wang}, \binits{Z.J.}},
\bauthor{\bsnm{Lee}, \binits{T.K.}}:
\batitle{Deep learning enhances polarization speckle for in vivo skin cancer detection}.
\bjtitle{Optics \& Laser Technology}
\bvolume{140},
\bfpage{107006}
(\byear{2021})
\end{barticle}
\endbibitem

\bibitem[\protect\citeauthoryear{Sezer and Sezer}{2020}]{sezer2020deep}
\begin{barticle}
\bauthor{\bsnm{Sezer}, \binits{A.}},
\bauthor{\bsnm{Sezer}, \binits{H.B.}}:
\batitle{Deep convolutional neural network-based automatic classification of neonatal hip ultrasound images: A novel data augmentation approach with speckle noise reduction}.
\bjtitle{Ultrasound in Medicine \& Biology}
\bvolume{46}(\bissue{3}),
\bfpage{735}--\blpage{749}
(\byear{2020})
\end{barticle}
\endbibitem

\bibitem[\protect\citeauthoryear{Gong et~al.}{2019}]{gong2019optical}
\begin{barticle}
\bauthor{\bsnm{Gong}, \binits{L.}},
\bauthor{\bsnm{Zhao}, \binits{Q.}},
\bauthor{\bsnm{Zhang}, \binits{H.}},
\bauthor{\bsnm{Hu}, \binits{X.-Y.}},
\bauthor{\bsnm{Huang}, \binits{K.}},
\bauthor{\bsnm{Yang}, \binits{J.-M.}},
\bauthor{\bsnm{Li}, \binits{Y.-M.}}:
\batitle{Optical orbital-angular-momentum-multiplexed data transmission under high scattering}.
\bjtitle{Light: Science \& Applications}
\bvolume{8}(\bissue{1}),
\bfpage{27}
(\byear{2019})
\end{barticle}
\endbibitem

\bibitem[\protect\citeauthoryear{Zhu et~al.}{2021}]{zhu2021compensation}
\begin{barticle}
\bauthor{\bsnm{Zhu}, \binits{Z.}},
\bauthor{\bsnm{Janasik}, \binits{M.}},
\bauthor{\bsnm{Fyffe}, \binits{A.}},
\bauthor{\bsnm{Hay}, \binits{D.}},
\bauthor{\bsnm{Zhou}, \binits{Y.}},
\bauthor{\bsnm{Kantor}, \binits{B.}},
\bauthor{\bsnm{Winder}, \binits{T.}},
\bauthor{\bsnm{Boyd}, \binits{R.W.}},
\bauthor{\bsnm{Leuchs}, \binits{G.}},
\bauthor{\bsnm{Shi}, \binits{Z.}}:
\batitle{Compensation-free high-dimensional free-space optical communication using turbulence-resilient vector beams}.
\bjtitle{Nature Communications}
\bvolume{12}(\bissue{1}),
\bfpage{1666}
(\byear{2021})
\end{barticle}
\endbibitem

\bibitem[\protect\citeauthoryear{Luo et~al.}{2022}]{luo2022computational}
\begin{barticle}
\bauthor{\bsnm{Luo}, \binits{Y.}},
\bauthor{\bsnm{Zhao}, \binits{Y.}},
\bauthor{\bsnm{Li}, \binits{J.}},
\bauthor{\bsnm{{\c{C}}etinta{\c{s}}}, \binits{E.}},
\bauthor{\bsnm{Rivenson}, \binits{Y.}},
\bauthor{\bsnm{Jarrahi}, \binits{M.}},
\bauthor{\bsnm{Ozcan}, \binits{A.}}:
\batitle{Computational imaging without a computer: Seeing through random diffusers at the speed of light}.
\bjtitle{eLight}
\bvolume{2}(\bissue{1}),
\bfpage{4}
(\byear{2022})
\end{barticle}
\endbibitem

\bibitem[\protect\citeauthoryear{LeCun et~al.}{2015}]{lecun2015deep}
\begin{barticle}
\bauthor{\bsnm{LeCun}, \binits{Y.}},
\bauthor{\bsnm{Bengio}, \binits{Y.}},
\bauthor{\bsnm{Hinton}, \binits{G.}}:
\batitle{Deep learning}.
\bjtitle{Nature}
\bvolume{521}(\bissue{7553}),
\bfpage{436}--\blpage{444}
(\byear{2015})
\end{barticle}
\endbibitem

\bibitem[\protect\citeauthoryear{Poggio et~al.}{2020}]{poggio2020theoretical}
\begin{barticle}
\bauthor{\bsnm{Poggio}, \binits{T.}},
\bauthor{\bsnm{Banburski}, \binits{A.}},
\bauthor{\bsnm{Liao}, \binits{Q.}}:
\batitle{Theoretical issues in deep networks}.
\bjtitle{Proceedings of the National Academy of Sciences}
\bvolume{117}(\bissue{48}),
\bfpage{30039}--\blpage{30045}
(\byear{2020})
\end{barticle}
\endbibitem

\bibitem[\protect\citeauthoryear{Zhuang et~al.}{2020}]{zhuang2020comprehensive}
\begin{barticle}
\bauthor{\bsnm{Zhuang}, \binits{F.}},
\bauthor{\bsnm{Qi}, \binits{Z.}},
\bauthor{\bsnm{Duan}, \binits{K.}},
\bauthor{\bsnm{Xi}, \binits{D.}},
\bauthor{\bsnm{Zhu}, \binits{Y.}},
\bauthor{\bsnm{Zhu}, \binits{H.}},
\bauthor{\bsnm{Xiong}, \binits{H.}},
\bauthor{\bsnm{He}, \binits{Q.}}:
\batitle{A comprehensive survey on transfer learning}.
\bjtitle{Proceedings of the IEEE}
\bvolume{109}(\bissue{1}),
\bfpage{43}--\blpage{76}
(\byear{2020})
\end{barticle}
\endbibitem

\bibitem[\protect\citeauthoryear{Van~Engelen and Hoos}{2020}]{van2020survey}
\begin{barticle}
\bauthor{\bsnm{Van~Engelen}, \binits{J.E.}},
\bauthor{\bsnm{Hoos}, \binits{H.H.}}:
\batitle{A survey on semi-supervised learning}.
\bjtitle{Machine Learning}
\bvolume{109}(\bissue{2}),
\bfpage{373}--\blpage{440}
(\byear{2020})
\end{barticle}
\endbibitem

\bibitem[\protect\citeauthoryear{Rahmani et~al.}{2020}]{rahmani2020actor}
\begin{barticle}
\bauthor{\bsnm{Rahmani}, \binits{B.}},
\bauthor{\bsnm{Loterie}, \binits{D.}},
\bauthor{\bsnm{Kakkava}, \binits{E.}},
\bauthor{\bsnm{Borhani}, \binits{N.}},
\bauthor{\bsnm{Te{\u{g}}in}, \binits{U.}},
\bauthor{\bsnm{Psaltis}, \binits{D.}},
\bauthor{\bsnm{Moser}, \binits{C.}}:
\batitle{Actor neural networks for the robust control of partially measured nonlinear systems showcased for image propagation through diffuse media}.
\bjtitle{Nature Machine Intelligence}
\bvolume{2}(\bissue{7}),
\bfpage{403}--\blpage{410}
(\byear{2020})
\end{barticle}
\endbibitem

\bibitem[\protect\citeauthoryear{Krenn et~al.}{2016}]{krenn2016twisted}
\begin{barticle}
\bauthor{\bsnm{Krenn}, \binits{M.}},
\bauthor{\bsnm{Handsteiner}, \binits{J.}},
\bauthor{\bsnm{Fink}, \binits{M.}},
\bauthor{\bsnm{Fickler}, \binits{R.}},
\bauthor{\bsnm{Ursin}, \binits{R.}},
\bauthor{\bsnm{Malik}, \binits{M.}},
\bauthor{\bsnm{Zeilinger}, \binits{A.}}:
\batitle{Twisted light transmission over 143 km}.
\bjtitle{Proceedings of the National Academy of Sciences}
\bvolume{113}(\bissue{48}),
\bfpage{13648}--\blpage{13653}
(\byear{2016})
\end{barticle}
\endbibitem

\bibitem[\protect\citeauthoryear{Wang et~al.}{2022}]{wang2022far}
\begin{barticle}
\bauthor{\bsnm{Wang}, \binits{F.}},
\bauthor{\bsnm{Wang}, \binits{C.}},
\bauthor{\bsnm{Chen}, \binits{M.}},
\bauthor{\bsnm{Gong}, \binits{W.}},
\bauthor{\bsnm{Zhang}, \binits{Y.}},
\bauthor{\bsnm{Han}, \binits{S.}},
\bauthor{\bsnm{Situ}, \binits{G.}}:
\batitle{Far-field super-resolution ghost imaging with a deep neural network constraint}.
\bjtitle{Light: Science \& Applications}
\bvolume{11}(\bissue{1}),
\bfpage{1}
(\byear{2022})
\end{barticle}
\endbibitem

\bibitem[\protect\citeauthoryear{Goy et~al.}{2019}]{goy2019high}
\begin{barticle}
\bauthor{\bsnm{Goy}, \binits{A.}},
\bauthor{\bsnm{Rughoobur}, \binits{G.}},
\bauthor{\bsnm{Li}, \binits{S.}},
\bauthor{\bsnm{Arthur}, \binits{K.}},
\bauthor{\bsnm{Akinwande}, \binits{A.I.}},
\bauthor{\bsnm{Barbastathis}, \binits{G.}}:
\batitle{High-resolution limited-angle phase tomography of dense layered objects using deep neural networks}.
\bjtitle{Proceedings of the National Academy of Sciences}
\bvolume{116}(\bissue{40}),
\bfpage{19848}--\blpage{19856}
(\byear{2019})
\end{barticle}
\endbibitem

\bibitem[\protect\citeauthoryear{Bender et~al.}{2021}]{bender2021circumventing}
\begin{barticle}
\bauthor{\bsnm{Bender}, \binits{N.}},
\bauthor{\bsnm{Sun}, \binits{M.}},
\bauthor{\bsnm{Y{\i}lmaz}, \binits{H.}},
\bauthor{\bsnm{Bewersdorf}, \binits{J.}},
\bauthor{\bsnm{Cao}, \binits{H.}}:
\batitle{Circumventing the optical diffraction limit with customized speckles}.
\bjtitle{Optica}
\bvolume{8}(\bissue{2}),
\bfpage{122}--\blpage{129}
(\byear{2021})
\end{barticle}
\endbibitem

\bibitem[\protect\citeauthoryear{Sagi and Rokach}{2018}]{sagi2018ensemble}
\begin{barticle}
\bauthor{\bsnm{Sagi}, \binits{O.}},
\bauthor{\bsnm{Rokach}, \binits{L.}}:
\batitle{Ensemble learning: A survey}.
\bjtitle{WIREs Data Mining and Knowledge Discovery}
\bvolume{8}(\bissue{4}),
\bfpage{1249}
(\byear{2018})
\end{barticle}
\endbibitem

\bibitem[\protect\citeauthoryear{Zhou}{2021}]{zhou2021ensemble}
\begin{bchapter}
\bauthor{\bsnm{Zhou}, \binits{Z.-H.}}:
\bctitle{Sensemble learning}.
In: \bbtitle{Machine Learning},
pp. \bfpage{181}--\blpage{210}.
\bpublisher{Springer},
\blocation{Singapore}
(\byear{2021})
\end{bchapter}
\endbibitem

\bibitem[\protect\citeauthoryear{Bilous et~al.}{2021}]{bilous2021handbook}
\begin{bbook}
\bauthor{\bsnm{Bilous}, \binits{R.}},
\bauthor{\bsnm{Donnelly}, \binits{R.}},
\bauthor{\bsnm{Idris}, \binits{I.}}:
\bbtitle{Handbook of Diabetes}.
\bpublisher{John Wiley \& Sons},
\blocation{Oxford}
(\byear{2021})
\end{bbook}
\endbibitem

\bibitem[\protect\citeauthoryear{Steiner et~al.}{2011}]{steiner2011optical}
\begin{barticle}
\bauthor{\bsnm{Steiner}, \binits{M.-S.}},
\bauthor{\bsnm{Duerkop}, \binits{A.}},
\bauthor{\bsnm{Wolfbeis}, \binits{O.S.}}:
\batitle{Optical methods for sensing glucose}.
\bjtitle{Chemical Society Reviews}
\bvolume{40}(\bissue{9}),
\bfpage{4805}--\blpage{4839}
(\byear{2011})
\end{barticle}
\endbibitem

\bibitem[\protect\citeauthoryear{Tuchin}{2008}]{tuchin2008handbook}
\begin{bbook}
\bauthor{\bsnm{Tuchin}, \binits{V.V.}}:
\bbtitle{Handbook of Optical Sensing of Glucose in Biological Fluids and Tissues}.
\bpublisher{CRC press},
\blocation{Boca Raton}
(\byear{2008})
\end{bbook}
\endbibitem

\bibitem[\protect\citeauthoryear{Ozana et~al.}{2014}]{ozana2014improved}
\begin{barticle}
\bauthor{\bsnm{Ozana}, \binits{N.}},
\bauthor{\bsnm{Arbel}, \binits{N.}},
\bauthor{\bsnm{Beiderman}, \binits{Y.}},
\bauthor{\bsnm{Mico}, \binits{V.}},
\bauthor{\bsnm{Sanz}, \binits{M.}},
\bauthor{\bsnm{Garcia}, \binits{J.}},
\bauthor{\bsnm{Anand}, \binits{A.}},
\bauthor{\bsnm{Javidi}, \binits{B.}},
\bauthor{\bsnm{Epstein}, \binits{Y.}},
\bauthor{\bsnm{Zalevsky}, \binits{Z.}}:
\batitle{Improved noncontact optical sensor for detection of glucose concentration and indication of dehydration level}.
\bjtitle{Biomedical optics express}
\bvolume{5}(\bissue{6}),
\bfpage{1926}--\blpage{1940}
(\byear{2014})
\end{barticle}
\endbibitem

\bibitem[\protect\citeauthoryear{Ozana et~al.}{2018}]{ozana2018remote}
\begin{barticle}
\bauthor{\bsnm{Ozana}, \binits{N.}},
\bauthor{\bsnm{Talman}, \binits{R.}},
\bauthor{\bsnm{Shemer}, \binits{A.}},
\bauthor{\bsnm{Schwartz}, \binits{A.}},
\bauthor{\bsnm{Polani}, \binits{S.}},
\bauthor{\bsnm{Califa}, \binits{R.}},
\bauthor{\bsnm{Beiderman}, \binits{Y.}},
\bauthor{\bsnm{Ruiz-Rivas}, \binits{J.}},
\bauthor{\bsnm{Garc}, \binits{J.}}:
\batitle{Remote photonic sensing of glucose concentration via analysis of time varied speckle patterns}.
\bjtitle{Advanced Materials Letters}
\bvolume{9}(\bissue{9}),
\bfpage{624}--\blpage{628}
(\byear{2018})
\end{barticle}
\endbibitem

\bibitem[\protect\citeauthoryear{Pal et~al.}{2022}]{pal2022non}
\begin{barticle}
\bauthor{\bsnm{Pal}, \binits{D.}},
\bauthor{\bsnm{Agadarov}, \binits{S.}},
\bauthor{\bsnm{Beiderman}, \binits{Y.}},
\bauthor{\bsnm{Beiderman}, \binits{Y.}},
\bauthor{\bsnm{Kumar}, \binits{A.}},
\bauthor{\bsnm{Zalevsky}, \binits{Z.}}:
\batitle{Non-invasive blood glucose sensing by machine learning of optic fiber-based speckle pattern variation}.
\bjtitle{Journal of Biomedical Optics}
\bvolume{27}(\bissue{9}),
\bfpage{097001}--\blpage{097001}
(\byear{2022})
\end{barticle}
\endbibitem

\bibitem[\protect\citeauthoryear{Gusev et~al.}{2020}]{gusev2020noninvasive}
\begin{barticle}
\bauthor{\bsnm{Gusev}, \binits{M.}},
\bauthor{\bsnm{Poposka}, \binits{L.}},
\bauthor{\bsnm{Spasevski}, \binits{G.}},
\bauthor{\bsnm{Kostoska}, \binits{M.}},
\bauthor{\bsnm{Koteska}, \binits{B.}},
\bauthor{\bsnm{Simjanoska}, \binits{M.}},
\bauthor{\bsnm{Ackovska}, \binits{N.}},
\bauthor{\bsnm{Stojmenski}, \binits{A.}},
\bauthor{\bsnm{Tasic}, \binits{J.}},
\bauthor{\bsnm{Trontelj}, \binits{J.}}:
\batitle{Noninvasive glucose measurement using machine learning and neural network methods and correlation with heart rate variability}.
\bjtitle{Journal of Sensors}
\bvolume{2020}(\bissue{1}),
\bfpage{9628281}
(\byear{2020})
\end{barticle}
\endbibitem

\bibitem[\protect\citeauthoryear{Pal et~al.}{2023}]{pal2023noninvasive}
\begin{barticle}
\bauthor{\bsnm{Pal}, \binits{D.}},
\bauthor{\bsnm{Kumar}, \binits{A.}},
\bauthor{\bsnm{Avraham}, \binits{N.}},
\bauthor{\bsnm{Eisenbach}, \binits{Y.}},
\bauthor{\bsnm{Beiderman}, \binits{Y.}},
\bauthor{\bsnm{Agdarov}, \binits{S.}},
\bauthor{\bsnm{Beiderman}, \binits{Y.}},
\bauthor{\bsnm{Zalevsky}, \binits{Z.}}:
\batitle{Noninvasive blood glucose sensing by secondary speckle pattern artificial intelligence analyses}.
\bjtitle{Journal of Biomedical Optics}
\bvolume{28}(\bissue{8}),
\bfpage{087001}--\blpage{087001}
(\byear{2023})
\end{barticle}
\endbibitem

\bibitem[\protect\citeauthoryear{Tang et~al.}{2020}]{tang2020non}
\begin{barticle}
\bauthor{\bsnm{Tang}, \binits{L.}},
\bauthor{\bsnm{Chang}, \binits{S.J.}},
\bauthor{\bsnm{Chen}, \binits{C.-J.}},
\bauthor{\bsnm{Liu}, \binits{J.-T.}}:
\batitle{Non-invasive blood glucose monitoring technology: A review}.
\bjtitle{Sensors}
\bvolume{20}(\bissue{23}),
\bfpage{6925}
(\byear{2020})
\end{barticle}
\endbibitem

\bibitem[\protect\citeauthoryear{Euliss and Sorensen}{1984}]{euliss1984dynamic}
\begin{barticle}
\bauthor{\bsnm{Euliss}, \binits{G.}},
\bauthor{\bsnm{Sorensen}, \binits{C.}}:
\batitle{Dynamic light scattering studies of concentration fluctuations in aqueous t-butyl alcohol solutions}.
\bjtitle{The Journal of chemical physics}
\bvolume{80}(\bissue{10}),
\bfpage{4767}--\blpage{4773}
(\byear{1984})
\end{barticle}
\endbibitem

\bibitem[\protect\citeauthoryear{Yoshizaki et~al.}{2003}]{yoshizaki2003dielectric}
\begin{barticle}
\bauthor{\bsnm{Yoshizaki}, \binits{K.}},
\bauthor{\bsnm{Urakawa}, \binits{O.}},
\bauthor{\bsnm{Adachi}, \binits{K.}}:
\batitle{Dielectric study of concentration fluctuation in solutions of polystyrene}.
\bjtitle{Macromolecules}
\bvolume{36}(\bissue{7}),
\bfpage{2349}--\blpage{2354}
(\byear{2003})
\end{barticle}
\endbibitem

\bibitem[\protect\citeauthoryear{Wu et~al.}{1991}]{wu1991enhanced}
\begin{barticle}
\bauthor{\bsnm{Wu}, \binits{X.-L.}},
\bauthor{\bsnm{Pine}, \binits{D.}},
\bauthor{\bsnm{Dixon}, \binits{P.}}:
\batitle{Enhanced concentration fluctuations in polymer solutions under shear flow}.
\bjtitle{Physical Review Letters}
\bvolume{66}(\bissue{18}),
\bfpage{2408}
(\byear{1991})
\end{barticle}
\endbibitem

\bibitem[\protect\citeauthoryear{Hecht}{2004}]{hecht2004city}
\begin{bbook}
\bauthor{\bsnm{Hecht}, \binits{J.}}:
\bbtitle{City of Light: The Story of Fiber Optics}.
\bpublisher{Oxford University Press, USA},
\blocation{New York}
(\byear{2004})
\end{bbook}
\endbibitem

\bibitem[\protect\citeauthoryear{Bozinovic et~al.}{2013}]{bozinovic2013terabit}
\begin{barticle}
\bauthor{\bsnm{Bozinovic}, \binits{N.}},
\bauthor{\bsnm{Yue}, \binits{Y.}},
\bauthor{\bsnm{Ren}, \binits{Y.}},
\bauthor{\bsnm{Tur}, \binits{M.}},
\bauthor{\bsnm{Kristensen}, \binits{P.}},
\bauthor{\bsnm{Huang}, \binits{H.}},
\bauthor{\bsnm{Willner}, \binits{A.E.}},
\bauthor{\bsnm{Ramachandran}, \binits{S.}}:
\batitle{Terabit-scale orbital angular momentum mode division multiplexing in fibers}.
\bjtitle{Science}
\bvolume{340}(\bissue{6140}),
\bfpage{1545}--\blpage{1548}
(\byear{2013})
\end{barticle}
\endbibitem

\bibitem[\protect\citeauthoryear{Willner}{2019}]{willner2019optical}
\begin{bbook}
\beditor{\bsnm{Willner}, \binits{A.}} (ed.):
\bbtitle{Optical Fiber Telecommunications}.
\bpublisher{Academic Press}, \blocation{???}
(\byear{2019})
\end{bbook}
\endbibitem

\bibitem[\protect\citeauthoryear{Wang and Isola}{2020}]{wang2020understanding}
\begin{bchapter}
\bauthor{\bsnm{Wang}, \binits{T.}},
\bauthor{\bsnm{Isola}, \binits{P.}}:
\bctitle{Understanding contrastive representation learning through alignment and uniformity on the hypersphere}.
In: \bbtitle{International Conference on Machine Learning},
pp. \bfpage{9929}--\blpage{9939}
(\byear{2020}).
\bcomment{PMLR}
\end{bchapter}
\endbibitem

\bibitem[\protect\citeauthoryear{Momeni et~al.}{2023}]{momeni2023backpropagation}
\begin{barticle}
\bauthor{\bsnm{Momeni}, \binits{A.}},
\bauthor{\bsnm{Rahmani}, \binits{B.}},
\bauthor{\bsnm{Mall{\'e}jac}, \binits{M.}},
\bauthor{\bsnm{Del~Hougne}, \binits{P.}},
\bauthor{\bsnm{Fleury}, \binits{R.}}:
\batitle{Backpropagation-free training of deep physical neural networks}.
\bjtitle{Science}
\bvolume{382}(\bissue{6676}),
\bfpage{1297}--\blpage{1303}
(\byear{2023})
\end{barticle}
\endbibitem

\bibitem[\protect\citeauthoryear{Ji et~al.}{2019}]{ji2019invariant}
\begin{bchapter}
\bauthor{\bsnm{Ji}, \binits{X.}},
\bauthor{\bsnm{Henriques}, \binits{J.F.}},
\bauthor{\bsnm{Vedaldi}, \binits{A.}}:
\bctitle{Invariant information clustering for unsupervised image classification and segmentation}.
In: \bbtitle{Proceedings of the IEEE/CVF International Conference on Computer Vision},
pp. \bfpage{9865}--\blpage{9874}
(\byear{2019})
\end{bchapter}
\endbibitem

\end{thebibliography}

\clearpage
\begin{appendices}
\captionsetup[figure]{labelfont={bf},labelformat={default},labelsep=period,name={Extended Data Fig.}}

\begin{figure}[htb]
\centering
\includegraphics[width=0.97\textwidth]{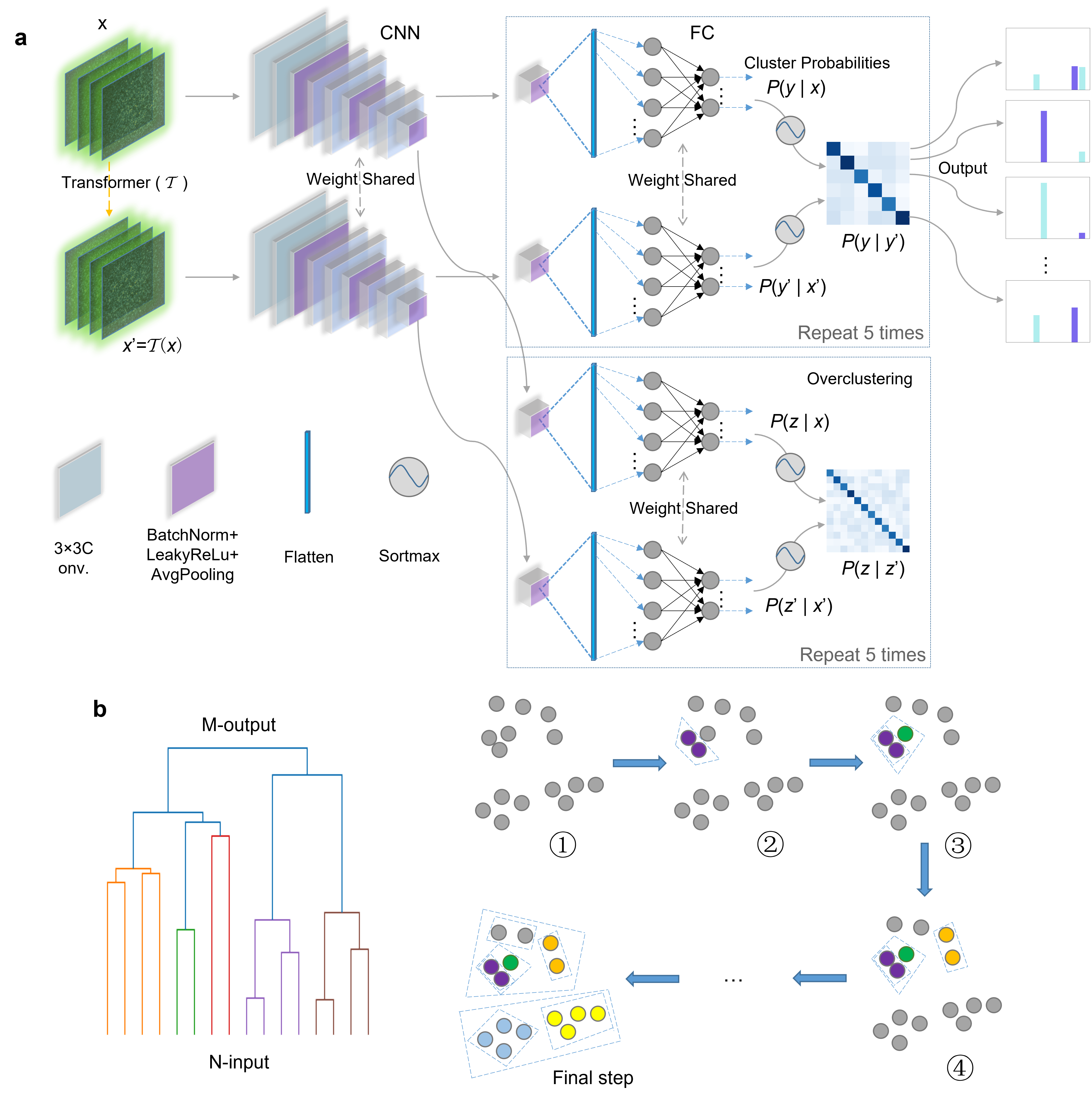}
\caption{Structures of SACN and SHACK. The input of SCAN is the speckle images and the output is the corresponding class probability, where the predicted result is the class with the maximum probability. The convolutional neural network (CNN) consists of six 3$\times$3 convolution layers and three composite layers, each containing BatchNormalization, Leaky ReLU and average pooling. The clustering and overclustering modules are a single-layer fully connected (FC) network but have significantly different numbers of output neurons. SHACK is based on the hierarchical nested clustering tree, which relies on computing the data similarity. Here, all $N$ speckles are regarded as independent unit, then the two speckles with the highest similarity are grouped together forming a new unit. This procedure is repeated until the preset number $M$ of classes is obtained, which means the achievement of clustering process.}\label{FigS1}
\end{figure} 

\clearpage
\begin{figure}[htb]
\centering
\includegraphics[width=0.7\textwidth]{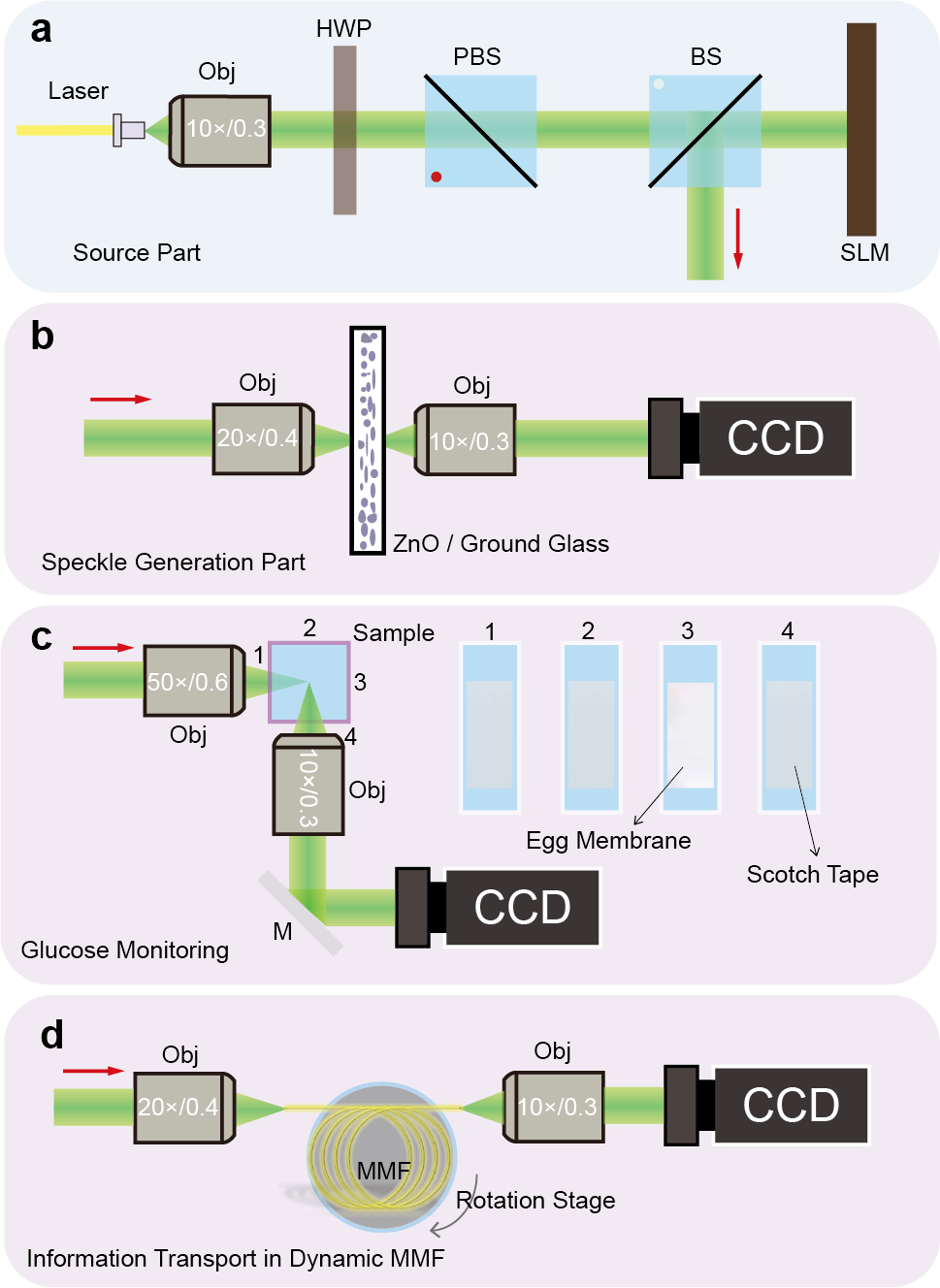}
\caption{Experimental setup. \textbf{a,} The light source and wavefront shaping part. This part is shared in all experiments. \textbf{b,} The part of speckle generation. The thick ZnO layer and ground glass are used to generate multiple- and single-layer scattering. \textbf{c,} The part of noninvasive glucose monitoring. The glucose solution is filled in the cuvette, with one surface wrapped in an egg membrane and the other three wrapped in Scotch tape. \textbf{d,} The part of information transport in dynamic MMF. The MMF is fixed in a rotation stage by two loose clamps. During the data transport, the rotation stage keeps moving, leading to a random disturbance in the MMF. The red arrows indicate the propagation direction of light. Obj: objective lens; HWP: half-wave plate; PBS: polarizing beam splitter; BS: beam splitter; SLM: spatial light modulator; CCD: charge-coupled device; M: mirror.}\label{FigS2}
\end{figure} 

\clearpage
\begin{figure}[htb]
\centering
\includegraphics[width=\textwidth]{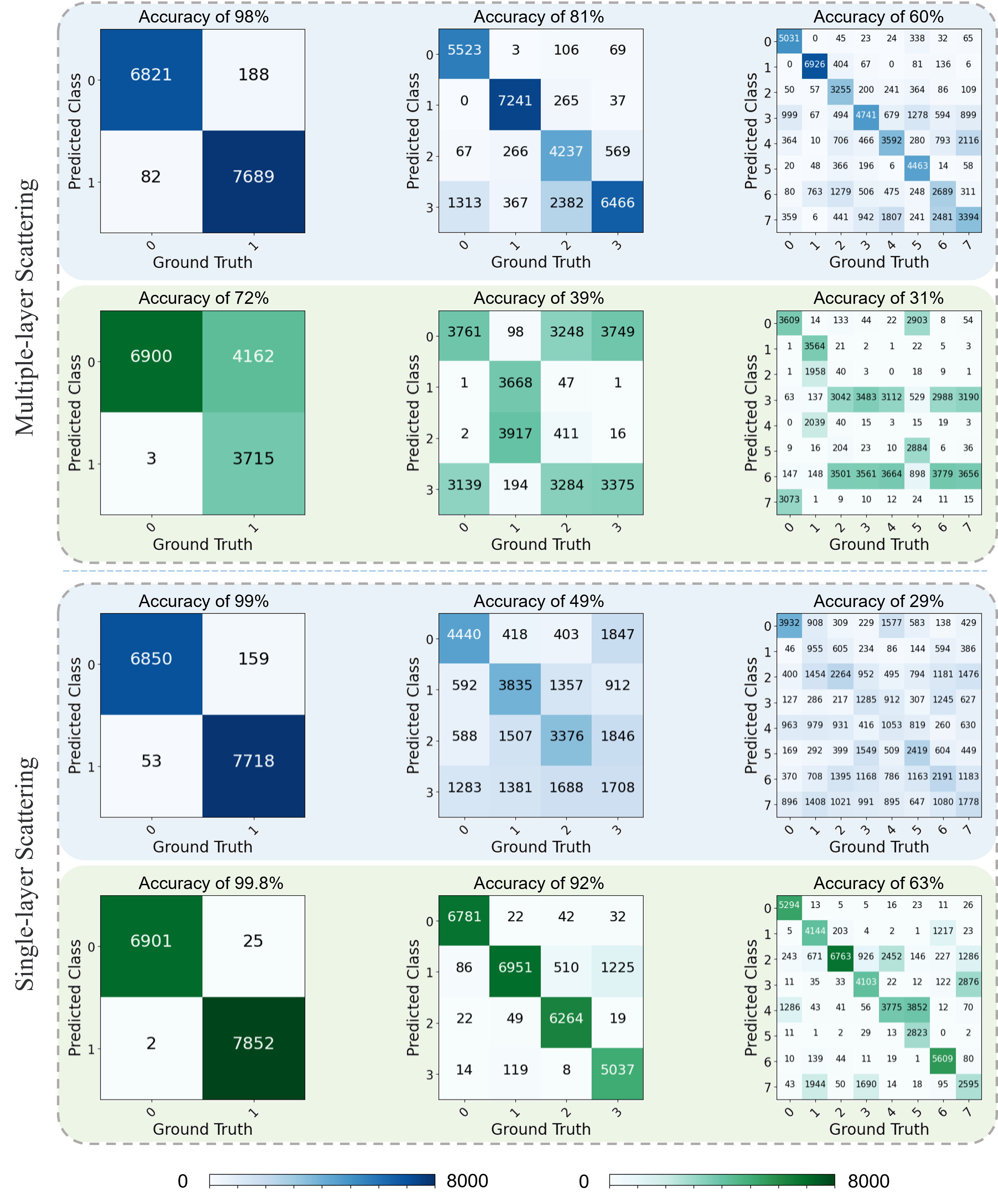}
\caption{Quantitative results for clustering different numbers of classes. The top and bottom panels show the results of clustering with speckles generated by multiple- and single-layer scattering, respectively. The green confusion matrices show the results predicted by SHACK, while the blue ones are generated by SCAN.}\label{FigS3}
\end{figure} 

\clearpage
\begin{figure}[htb]
\centering
\includegraphics[width=\textwidth]{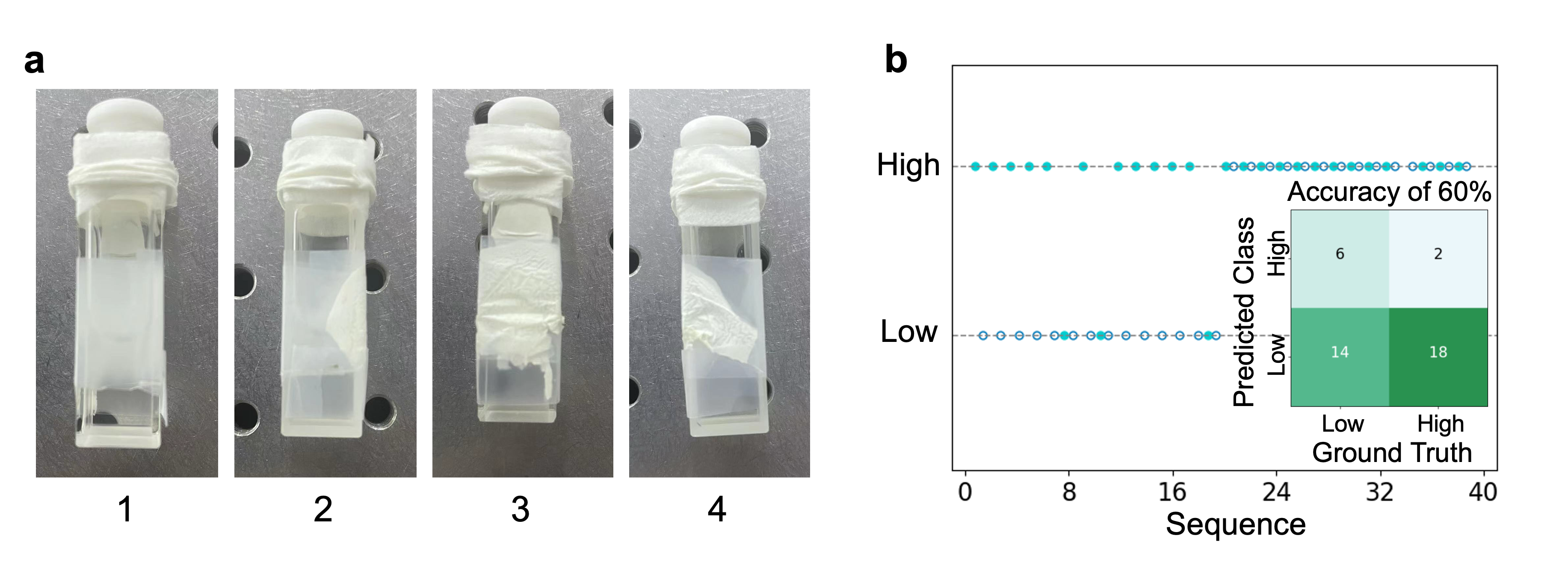}
\caption{The cuvette of glucose solution and control experiment. \textbf{a,} The four surfaces of the actual cuvette. \textbf{b,} The results of the control group with clear water. Here, the glucose solution is replaced by clear water to prove that SURE achieves the clustering according to the concentration rather than the different timestamp.}\label{FigS4}
\end{figure} 

\clearpage
\begin{figure}[htb]
\centering
\includegraphics[width=\textwidth]{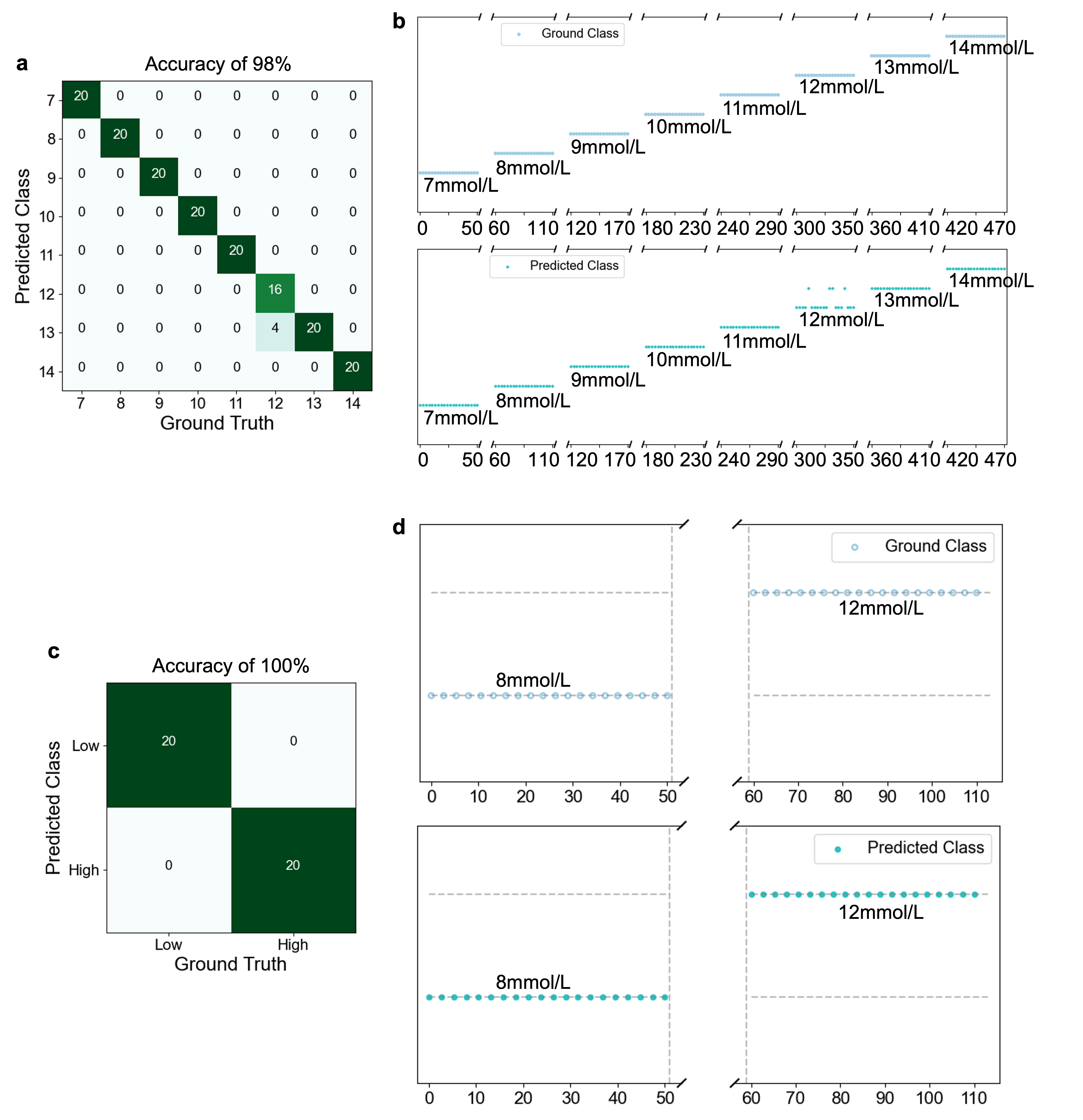}
\caption{Quantitative results for the glucose solution with different concentration levels. \textbf{a} and \textbf{b} are the clustering results with 8 concentration levels. \textbf{c} and \textbf{d}  are the same as Fig. \ref{Fig3}e but the initial 3mL of water is replaced by a 17:3 water-to-milk ratio to mimic a liquid environment with complex scattering.}\label{FigS5}
\end{figure} 

\clearpage
\begin{figure}[htb]
\centering
\includegraphics[width=\textwidth]{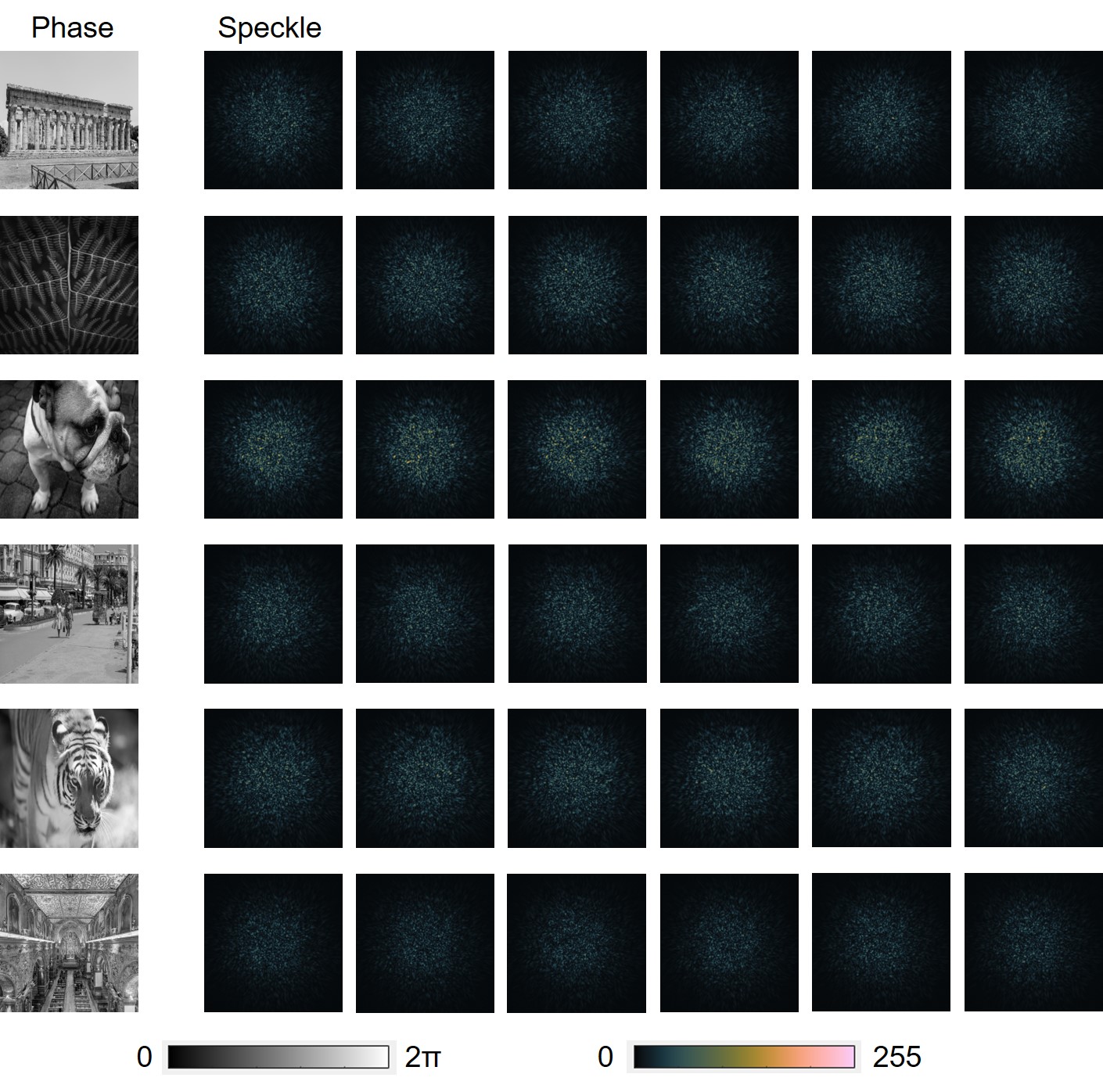}
\caption{Some raw data in the information transport in dynamic MMF. We randomly select six images from the DIV2K dataset and show six speckles corresponding to the phase images collected at different times. Due to the random disturbance in the MMF, these speckles arising from identical phase modulation are significantly different.}\label{FigS6}
\end{figure} 

\clearpage
\begin{figure}[htb]
\centering
\includegraphics[width=\textwidth]{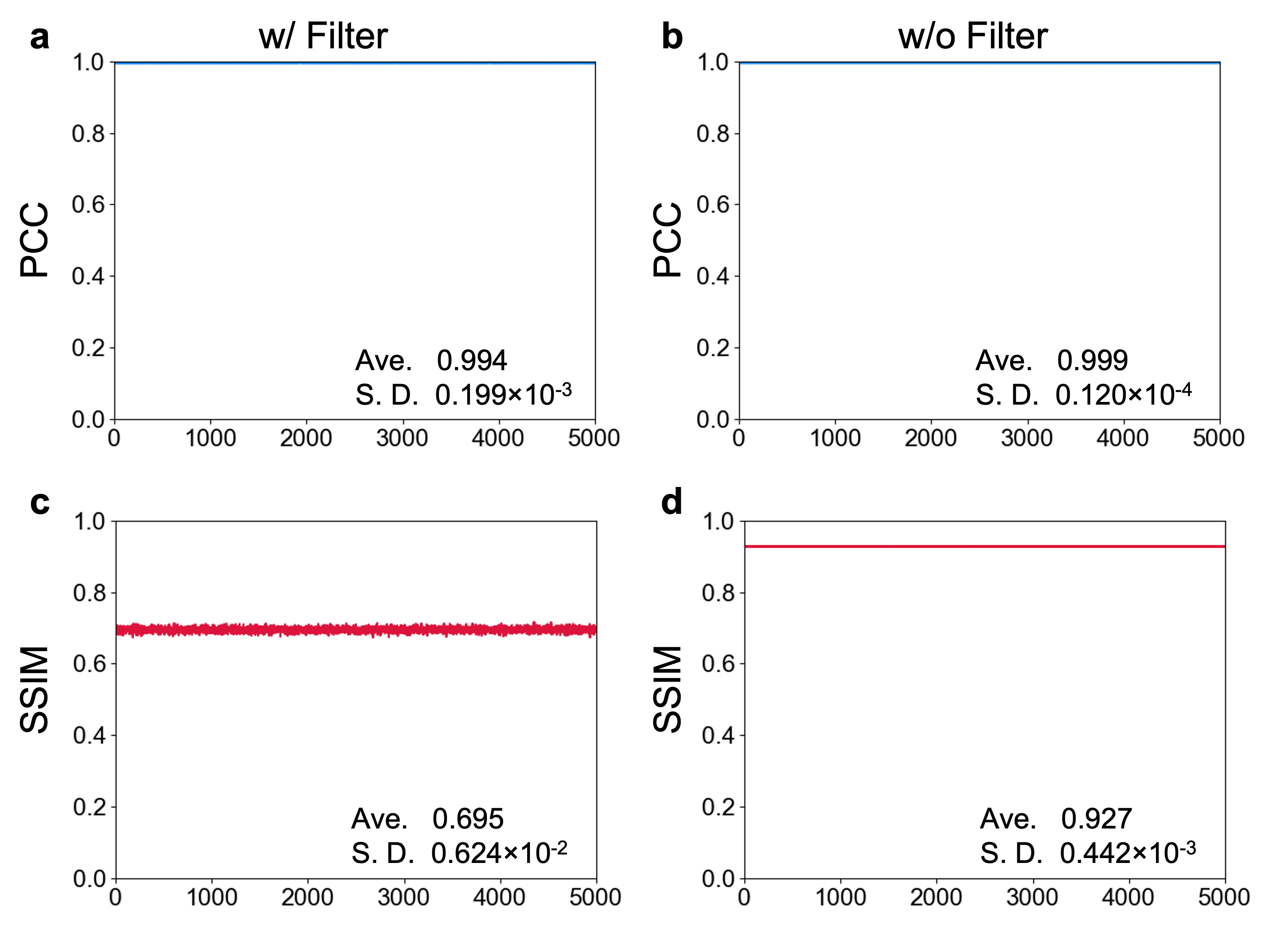}
\caption{The results for the data transport in dynamic MMF. The image in Fig. \ref{Fig4}d is repeatedly transmitted 5000 times to obtain statistics of transport fidelity. \textbf{a} and \textbf{b} employ the Pearson correlation coefficient (PCC), while \textbf{c} and \textbf{d} use the structural similarity index measure (SSIM) to evaluate image fidelity. Ave. stands for the average value, and S.D. stands for the standard deviation.}\label{FigS7}
\end{figure} 




\end{appendices}

\end{document}